\documentclass[aps,reprint,twocolumn,superscriptaddress,10pt]{revtex4-1}
\usepackage[dvipdfmx]{graphicx}
\usepackage{siunitx}
\usepackage{amsfonts}
\usepackage{amsmath}
\usepackage{amssymb}
\usepackage{bm}
\usepackage{graphicx}
\usepackage{dcolumn}
\usepackage{mciteplus}
\usepackage{euscript}
\usepackage{multirow}
\usepackage{color}
\usepackage{textcomp}
\usepackage{gensymb}
\usepackage{dblfloatfix}
\usepackage{float}
\usepackage{enumitem}
\usepackage{xcolor}
\usepackage{sepfootnotes}

\graphicspath{{./}{./figs/}{./figs_supplemental/}}

\newcommand{\rfig}[1]{Fig.~\ref{#1}}
\newcommand{\rFig}[1]{Figure~\ref{#1}}
\newcommand{\rtbl}[1]{Table~\ref{#1}}

\newcommand{\qsgw}{\rm{QSGW}}
\newcommand{\ef}{E_{\rm F}}
\newcommand{\eg}{E_{\rm g}}

\newcommand\etal{\textit{et al.}}

\newcommand{\vxc}{V^{\rm xc}}

\newcommand{\cgt}{\text{Cr}\text{Ge}\text{Te}_3}
\newcommand{\vvi}{\text{VI}_3}
\newcommand{\cri}{\text{CrI}_3}
\newcommand{\fgt}{\text{Fe}_3\text{GeTe}_2}

\begin{document}
\title{Role of nonlocality in exchange-correlation for magnetic two-dimensional van der Waals  materials}
\author{Y. Lee}
\affiliation{Ames Laboratory, U.S.~Department of Energy, Ames, Iowa 50011, USA}
\author{Takao Kotani}
\affiliation{Department of Applied Mathematics and Physics, Tottori University, Tottori 680-8552, Japan} 
\author{Liqin Ke}
\email[Corresponding author: ]{liqinke@ameslab.gov}
\affiliation{Ames Laboratory, U.S.~Department of Energy, Ames, Iowa 50011, USA}
\date{\today} 

\begin{abstract}

To obtain accurate independent-particle descriptions for ferromagnetic two-dimensional van der Waals materials, we apply the quasiparticle self-consistent $GW$ (QSGW) method to VI$_3$, CrI$_3$, CrGeTe$_3$, and Fe$_3$GeTe$_2$.
QSGW provides a description of the nonlocal exchange-correlation term in the one-particle Hamiltonian.
The nonlocal term is important not only as the $U$ of density functional theory (DFT)+$U$ but also for differentiating occupied and unoccupied states in semiconductors.
We show the limitations of DFT+$U$ in mimicking QSGW.


\end{abstract}

\maketitle	

\paragraph*{Introduction.---}
The recent experimental realization of magnetic two-dimensional (2D) van der Waals (vdW) materials has generated great interest for exploiting novel 2D magnetism
and for applications such as energy-efficient ultracompact spin-based electronics~\cite{zhong2017sa}.
Long-range ferromagnetic ordering in the atomically thin systems was first demonstrated in the $\cgt$ bilayer~\cite{gong2017n} and $\cri$ monolayer~\cite{huang2017n}, albeit only at very low temperatures.
Later, Deng $\etal$~\cite{deng2018n} showed that an electric field could drastically increase the Curie temperature, $T_\text{C}$, of a $\fgt$ monolayer up to room temperature.
Recently, $\vvi$ has been identified as the first vdW hard ferromagnet with a large coercivity~\cite{tian2019jacs,he2016jmcc,kong2019am}.
Spurred by these experiments, many theoretical efforts have been published treating
magnetic 2D vdW materials (m2Dv)~\cite{baidya2018prb,fang2018prb,jiang2018nl,kulish2017jmcc,sivadas2018nl,menichetti20192m,torelli20182m,lado20172m}.

We are also witnessing the recent revolutionary development of materials informatics (MI).
For example, Mounet $\etal$~\cite{mounet2018nn}
have employed a computational MI to search for 2D exfoliable materials by multi-level screening from the databases of experimentally known compounds.
The quality of such work largely depends on the choice of the first-principles method used for the final screening.
In the future, such an MI procedure may be applied to m2Dv.
Then the first-principles method used in MI should be as reliable as possible and with no adjustable parameters for each material.

Until now, m2Dv has been theoretically treated mostly within density functional theory (DFT)+$U$, with a single Hubbard $U$ applied on the cation-$3d$ orbitals, as in Refs.[\onlinecite{jang2019prm,hao2018sb, li2014jmcc,son2019prb}].
Phenomenological theories, such as DFT+$U$ and dynamical mean-field theory, are very useful for various material systems.
However, it is not clear that one can use DFT+$U$ for the above-mentioned MI, because of the limitation of the single parameter $U$, as we illustrate in the following.

First, the cation-$3d$ bands in m2Dv contain more degrees of freedom than a single $U$ parameter can describe.
Although DFT+$U$  may adjust overall splitting between occupied and unoccupied $3d$ bands for each spin, it ignores the $k$ dependence and frequency dependence of effective interactions and thus the interaction anisotropy regarding in-plane and out-of-plane $3d$ orbitals in m2Dv can not be adequately treated.
An idea using many parameters for the $U$ term would be hard to implement because of the difficulty in determining the unique parameters.

Second, the relative positions of cation-$3d$ and anion-$p$ bands are not directly controlled by onsite $U$.
For example, even in nonmagnetic CdO where we expect no $U$ effect because Cd-$4d$ states are fully occupied, we see the center of occupied $4d$ states can be pushed down about \SI{2}{eV}(see Fig.~A1 in Ref.~[\onlinecite{deguchi2016jjap}]) in comparison with DFT.                                 
Note that the relative positions of and hybridizations between cation-$3d$ and anion-$p$ can be important to determine the super-exchange coupling in m2Dv.

In this Rapid Communication, we apply the quasiparticle self-consistent $GW$ ($\qsgw$) method~\cite{kotani2007prb,kotani2014jpsj,deguchi2016jjap} to m2Dv, including $\vvi$, $\cri$, $\cgt$, and $\fgt$.
$\qsgw$ has been applied to a wide range of materials and shown to be the most reliable method available to determine the one-particle Hamiltonian $H_0$, which defines the independent-particle picture of a particular material.
For each material, an accurate $H_0$ is the key to evaluate all of its physical quantities theoretically.
We will show that $\qsgw$ reasonably describes electronic structures consistent with experiments for all m2Dv treated here.
Then we will examine whether DFT+$U$ can mimic the band structures obtained in $\qsgw$.
We will demonstrate the serious limitations of DFT in treating m2Dv, corresponding to the two reasons discussed above.

\paragraph*{Methods.---}
First, let us recall the $GW$ approximation (GWA)~\cite{hedin1969book,aryasetiawan1998ropi}.
GWA can be applied to any one-particle Hamiltonian $H_0$, for example, to the Kohn-Sham Hamiltonian of DFT.
In GWA, we calculate the self-energy $\Sigma({\bf r},{\bf r}',\omega)=\Sigma(1,2) =iG_0(1,2)W(1+,2)$.
Here $G_0=1/(\omega -H_0)$ is the Green's function of $H_0$; $W$ is the
dynamically screened Coulomb interaction calculated using $G_0$, usually in the random phase approximation (RPA). 
Then we can determine the quasiparticle energies with $\Sigma({\bf r},{\bf r}',\omega)$ in the place of the exchange-correlation (xc) potential. The reliability of this one-shot method, so-called $G_0W_0$, depends on the reliability of $H_0$.

Thus, the main theoretical problem of $G_0W_0$ is how to determine $H_0$ to which we apply GWA.
For this purpose, various self-consistent schemes have been developed.
In practice, a partial self-consistency is often employed due to the demanding nature of computation or the intrinsic problems of the methods~\cite{kutepov2016prb}.
In the so-called energy-only self-consistent $GW$ method~\cite{hybertsen1986prb,shishkin2007prl}, the eigenfunctions are fixed while only the one-particle energies are iterated to reach consistency.
In a $GW_0$ method~\cite{shishkin2007prl}, one may calculate $W$ using DFT $G_0$, but iterate $G$ untill convergence.

$\qsgw$~\cite{kotani2007prb,kotani2014jpsj,sakakibara2020prb} is given as a self-consistent perturbation method based on the quasiparticle picture within GWA.
The full many-body Hamiltonian $H$ is divided into $H=H_0+(H-H_0)$, then $(H-H_0)$ is treated as a perturbation in GWA.
The self-consistent perturbation requires that we should determine $H_0$ so that the term generated in GWA due to $(H-H_0)$ gives virtually zero.

Based on this idea, we generate the $\qsgw$ xc potential $V^{\rm xc}_{\rm QSGW}$ from the self-energy $\Sigma({\bf r},{\bf r}',\omega)$ obtained in GWA as
\begin{equation}
V^{\rm xc}_{\rm QSGW} = \frac{1}{2}\sum_{ij} |\psi_i\rangle 
       \left\{ {{\rm Re}[\Sigma(\epsilon_i)]_{ij}+{\rm Re}[\Sigma(\epsilon_j)]_{ij}} \right\}
       \langle\psi_j|.  
\label{eq:vxc}
\end{equation}
Here $\epsilon_i$ and $|\psi_i\rangle$ are eigenvalues and eigenfunctions, respectively, of Hamiltonian $H_0$.
Re denotes the Hermitian part.
$\Sigma_{ij}(\omega)=\langle \psi_i|\Sigma(\omega)|\psi_j \rangle =\int d^3{\bf r} \int d^3{\bf r}' \psi_i^*({\bf r}) \Sigma({{\bf r}},{{\bf r}'},\omega) \psi_j({\bf r}')$.
With Eq.(\ref{eq:vxc}), we have a mapping to generate a new $H_0$, $H_0^{(i)} \to H_0^{(i+1)}$.
This is repeated until $H_0$ is converged.
Note that $G_0W_0$ applied to this self-consistent $H_0$ does not cause corrections of the quasiparticle energies because of this self-consistency.

$\qsgw$, as it is, tends to systematically overestimate exchange effects, especially for bandgaps~\cite{deguchi2016jjap,van-schilfgaarde2006prl,kotani2007prb}.
This can be due to the underestimation of the screening effect in RPA, which neglects electron-hole correlations in the proper polarization function~\cite{van-schilfgaarde2006prl,shishkin2007prl}, and/or the neglect of the screening effect of phonons~\cite{botti2013prl}.
Shishkin $\etal$~\cite{shishkin2007prl} performed calculations that include the correlation via the vertex correction for $W$ and demonstrated very reliable predictions of band gaps by recovering the screening underestimation.
However, their methods are too computationally demanding to apply to the materials treated here.
Based on the observation that the underestimations are rather systematic in various systems~\cite{bhandari2018prm}, we here use a hybrid $\qsgw$ method, $\qsgw$80~\cite{chantis2006prl,deguchi2016jjap}, which uses an empirical mixing of $\vxc=0.8\vxc_{\rm QSGW}+0.2V^{\rm xc}_{\rm LDA}$.
$\qsgw$80 is taken to be a substitution of the method by Shishkin $\etal$ to remedy the underestimation quickly and efficiently.
Unless specified, all $\qsgw$ calculations in this work are carried out in $\qsgw$80, referred to hereafter as $\qsgw$, for simplicity.

The nonlocality of $V^{\rm xc}_{\rm QSGW}$ provides a natural description of the correct independent-particle picture.
Generally speaking, we can classify this nonlocality into two parts: on site and off site.
The on-site nonlocality, which can differentiate five $3d$ orbitals, can be approximated, to a certain extent, by the Hubbard $U$ in DFT$+U$.
The off-site nonlocality is critical to generate bandgaps in semiconductors.
To illustrate this, let us consider a hydrogen dimer H$_2$.
To lower the highest occupied molecular orbital (HOMO) energy without changing the shape of eigenfunctions, one needs to introduce a projector of HOMO.
The corresponding projector is naturally represented by a nonlocal potential, and the screened exchange contribution in $V^{\rm xc}_{\rm QSGW}$ works exactly as such a projector.

Furthermore, in contrast to the hybrid functional methods, the electron screening effects on the exchange is calculated explicitly in $\qsgw$.
The screened Coulomb interaction $W$, which determines the screened exchange, is spatially dependent and self-consistently determined without any system-dependent parameters.
On the other hand, in the hybrid functional methods such as Heyd-Scuseria-Ernzerhof (HSE), the xc functional is obtained by mixing the DFT xc with the Hartree-Fock (HF) exchange, which is calculated using the bare Coulomb interaction kernel.
The mixing parameter solely mimics the screening effect.
This limits the universality of the hybrid functional methods because the screening effects vary significantly between metals and semiconductors, and their spatial dependence could be important in anisotropic systems, which can be hard to be mimicked by one single parameter.
In fact, He and Franchini~\cite{he2019prb} showed that the mixing could be very material-dependent.
Thus, the explicit treatment of screened exchange allows $\qsgw$ to treat complex subjects such as metal/insulator interfaces, and also m2Dv, which contain both features of semiconductor and anisotropic magnetic materials.

\paragraph*{Computational details.---}
We use the $\qsgw$ method from the \textsc{ecalj} package~\cite{kotani2014jpsj}, which is implemented with a mixed basis and allows automatic interpolation of self-energy in the whole Brillouin zone without resorting to the \textsc{wannier90} techniques~\cite{mostofi2008cpc,mostofi2014cpc}.
The spin-orbit coupling (SOC) is included as a perturbation~\cite{deguchi2016jjap} after we attain the self-consistency of $\qsgw$.
We employed the experimental lattice parameters~\cite{tian2019jacs,mcguire2015cm,carteaux1995jpcm,deiseroth2006ejic} for calculations.
As for DFT+$U$, we use both fully-localized-limit (FLL)~\cite{liechtenstein1995prb} and around-the-mean-field (AMF)~\cite{petukhov2003prb} double-counting schemes to investigate the dependence of band structures on the correlation parameter $U$, which is applied on the cation-$3d$ orbitals.
All DFT and DFT+$U$ calculations are carried out within the generalized gradient approximation using the functional of Perdew, Burke, and Ernzerhof (PBE)~\cite{perdew1996prl}.

\begin{figure}[htb]
	\centering
	\begin{tabular}{c}
	\includegraphics[width=.98\linewidth,clip]{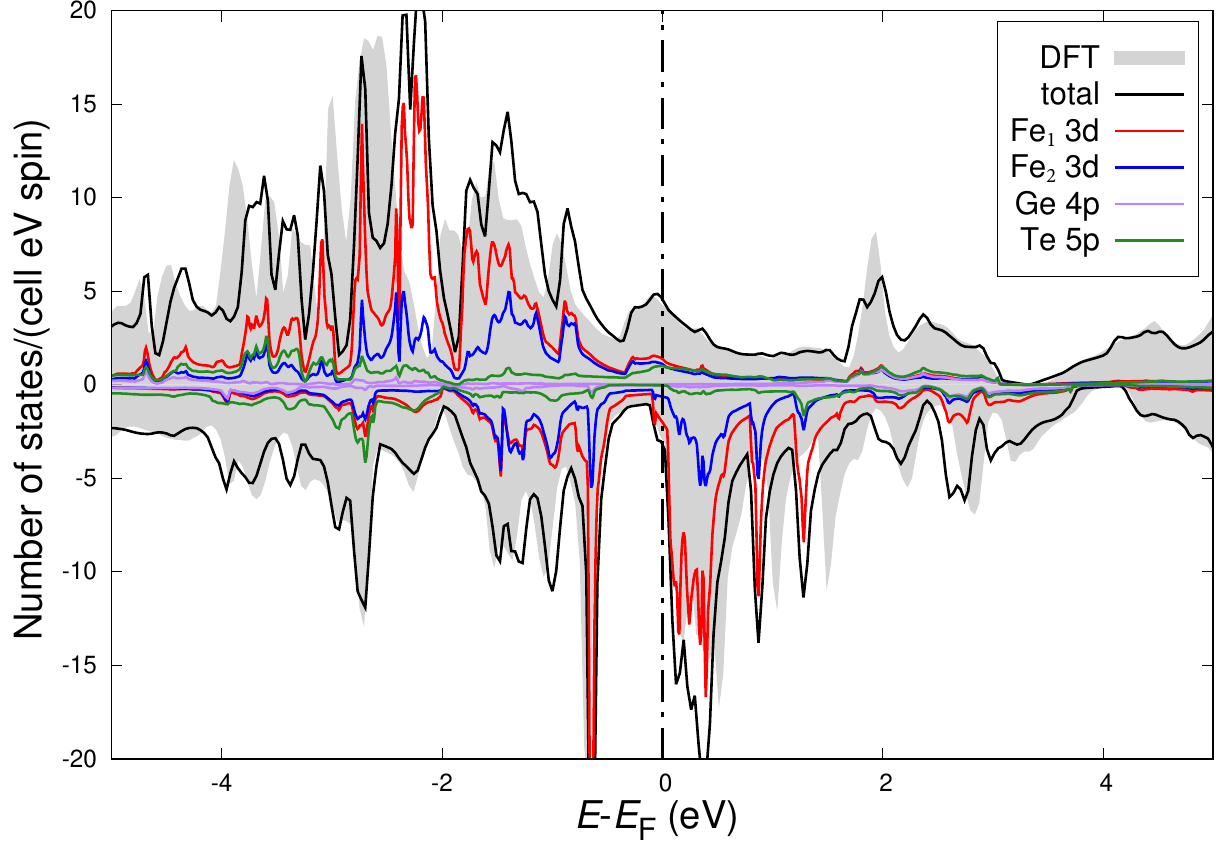} 
	\end{tabular}
	\caption{Total and atom-resolved partial density of states calculated using $\qsgw$ in $\fgt$. For comparison, DOS obtained by DFT is shown (shaded area). Spin-orbit coupling is not included.}
	\label{fig:dos_fgt}
\end{figure}

\paragraph*{Results.---}
$\fgt$ is a metallic m2Dv and has a higher $T_\text{C}$ than its semiconducting counterparts~\cite{deng2018n}.
\rFig{fig:dos_fgt} shows the  total density of states (DOS) and partial density of states (PDOS) calculated in $\qsgw$.
DOS obtained by DFT is also shown for comparison.
Both $\qsgw$ and DFT suggest that $\fgt$ is a metal, as found in experiments.
DOS are dominated by Fe-$3d$ states in this energy window. 
The Fermi level $\ef$ is located at a pseudogap of Fe$_1$-$3d$ states in the minority-spin channel.
$\qsgw$ gives slightly narrower $3d$ bands than DFT, suggesting a somewhat stronger localization of electron states in $\qsgw$.
Indeed, such $3d$-band narrowing is rather general in $\qsgw$ as shown in Refs.~[\onlinecite{kotani2009jpcm,jang2015sr}].
Considering the fact that $\qsgw$ describes metals such as bcc Fe and also transition-metal oxides such as NiO very well, our result supports the applicability of DFT to $\fgt$.
For a band structure comparison between DFT and QSGW, see the Supplemental Material~\cite{sm}.

Note the difficulty of hybrid functionals such as HSE applied to m2Dv without a choice of material-dependent parameters.
For example, one usually uses a mixing parameter $a=0.25$ for semiconductors.
However, it was found that $a=0.15$~\cite{patanachai2014jctc} is optimum for transition-metal oxides.
Furthermore, $a=0$ is apparently good for bcc Fe while HSE06 gives a magnetic moment of 2.89 $\mu_\text{B}$/Fe~\cite{yu2016jctc}.
Since semiconducting and metallic features coexist with transition metals in m2Dv, we can hardly expect HSE to work well for m2Dv.
We think that $\qsgw$ is the optimal choice to describe electronic structures of m2Dv along the line of MI.

\begin{table}[hbt]
 \caption{Bandgaps $\eg$ (\si{eV}) calculated in DFT and $\qsgw$, with and without SOC.
   Experimental values are listed to compare.
   The reported theoretical $\eg$ are in the range of 0.74--1.6, 0--0.43, and 0--1.0~\si{eV} for bulk $\cri$~\cite{wang2011jpcm, jiang2018nl}, $\cgt$~\cite{menichetti20192m, fang2018prb}, and $\vvi$~\cite{he2016jmcc, son2019prb} respectively.
}
 \sepfootnotecontent {a}{Resistivity measurement: \SI{0.32}{eV}~\cite{son2019prb};
   optical reflectance: \SI{0.6}{eV}~\cite{kong2019am};
   optical transmittance: \SI{0.67}{eV}~\cite{son2019prb}.
 }
 \sepfootnotecontent {b}{Optical transition measurement~\cite{dillon1965jap}.}
 \sepfootnotecontent {c}{Angle-resolved photoemission spectroscopy (ARPES) measurements: \SI{0.38}{eV}~\cite{li2018prb} and \SI{0.2}{eV}~\cite{suzuki2019prb}; resistivity measurement: \SI{0.2}{eV}~\cite{ji2013jap}; scanning tunneling microscopy (STM) measurement: \SI{0.74}{eV}~\cite{hao2018sb}.}
 
  \label{tbl:gap}%
        \def\arraystretch{1.2}
        \bgroup
        \begin{tabular*}{\linewidth}{l @{\extracolsep{\fill}} lllllll}
          \hline          
          \hline
          Compound &   & Experiment & DFT  & $\qsgw$ & DFT & $\qsgw$   \\
                   &   &            & SOC  &  SOC    & &               \\
          \hline
          $\vvi$   & & 0.32--0.67\sepfootnote{a}    & 0     & 0.53   & 0 &0.75 \\          
          $\cri$   & & 1.2 \sepfootnote{b}          & 0.78 & 1.68 & 1.07& 2.23 \\
          $\cgt$   & & 0.20--0.74\sepfootnote{c}   & 0.19 & 0.66 & 0.42& 0.99 \\
          \hline
          \hline          
        \end{tabular*}
        \egroup
\end{table}

\rtbl{tbl:gap} summarizes the experimental and our calculated $\eg$ values in m2Dv.
Unlike DFT, $\qsgw$ correctly predicts $\vvi$ as a semiconductor.
It is worth noting that $G_0W_0$ applied to DFT does not open the gap in $\vvi$, as it does in VO$_2$, demonstrating the necessity of self-consistency of $GW$ calculations as in $\qsgw$.
For $\cgt$, $\qsgw$ gives $\eg=\SI{0.66}{\eV}$, within the range of reported experimental values of \SIrange{0.20}{0.74}{eV}, while DFT gives a much smaller value of $\eg=\SI{0.19}{eV}$.
On the other hand, in $\cri$, $\qsgw$ gives $\eg=\SI{1.68}{eV}$, 35\% larger than the only reported experimental value of \SI{1.2}{eV}.
This difference is somewhat larger than expected, considering that $\qsgw$ produces $\eg$ within $\sim10\%$ difference for a wide range of materials~\cite{deguchi2016jjap}.


SOC reduces the calculated $\eg$ in all three semiconducting compounds, as shown in ~\rtbl{tbl:gap}, especially within $\qsgw$.
The strong SOC effects on $\eg$ are due to the heavy anion atoms in the compounds.
I- and Te-$5p$ orbitals have rather large SOC constants, $\xi_p=\SIrange{0.9}{1.0}{eV}$, while V- and Cr-$3d$ orbitals have $\xi_d=\SIrange{20}{30}{meV}$.
The contribution of SOC to $\eg$ of $\cri$ in $\qsgw$ (\SI{0.55}{eV}) is about twice as large as in DFT (\SI{0.29}{eV}).

\begin{figure}[htb]
	\centering
	\begin{tabular}{c}
          \includegraphics[width=.98\linewidth,clip]{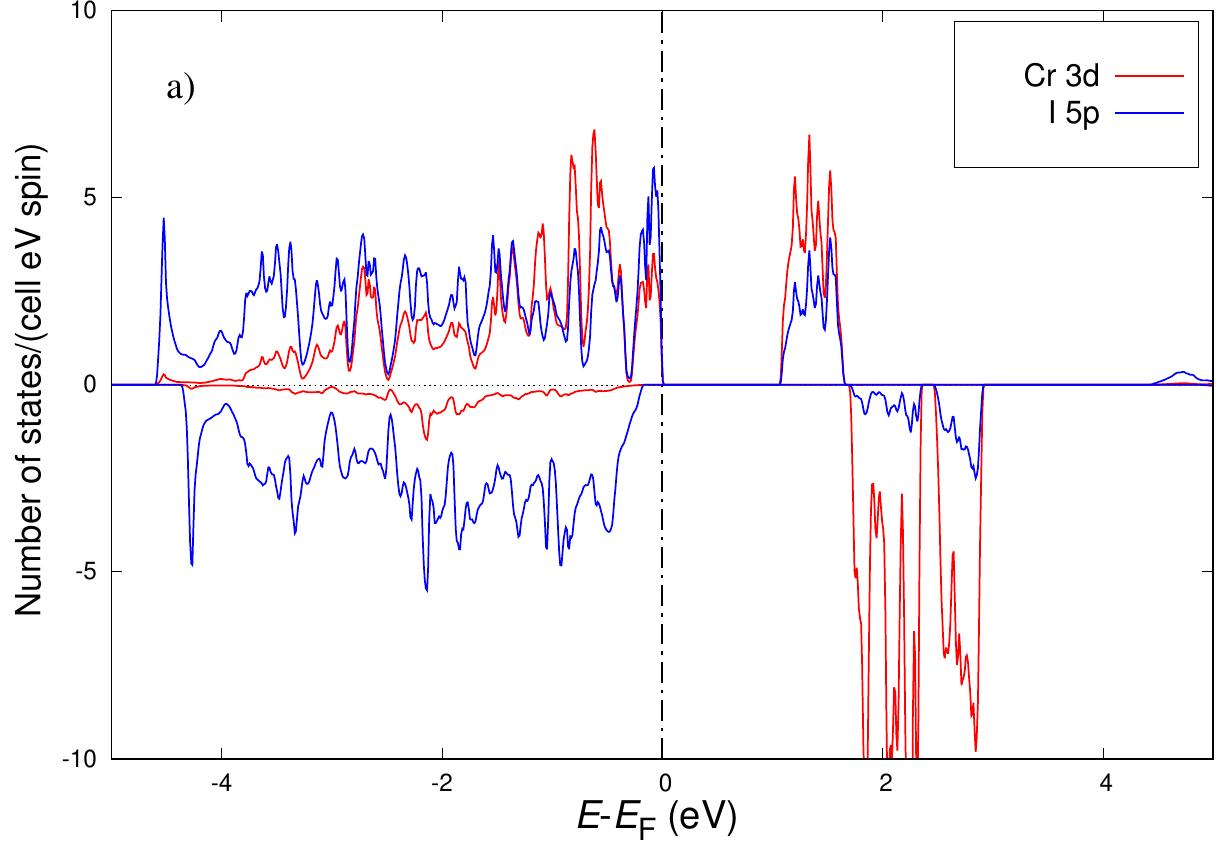} \\
          \includegraphics[width=.98\linewidth,clip]{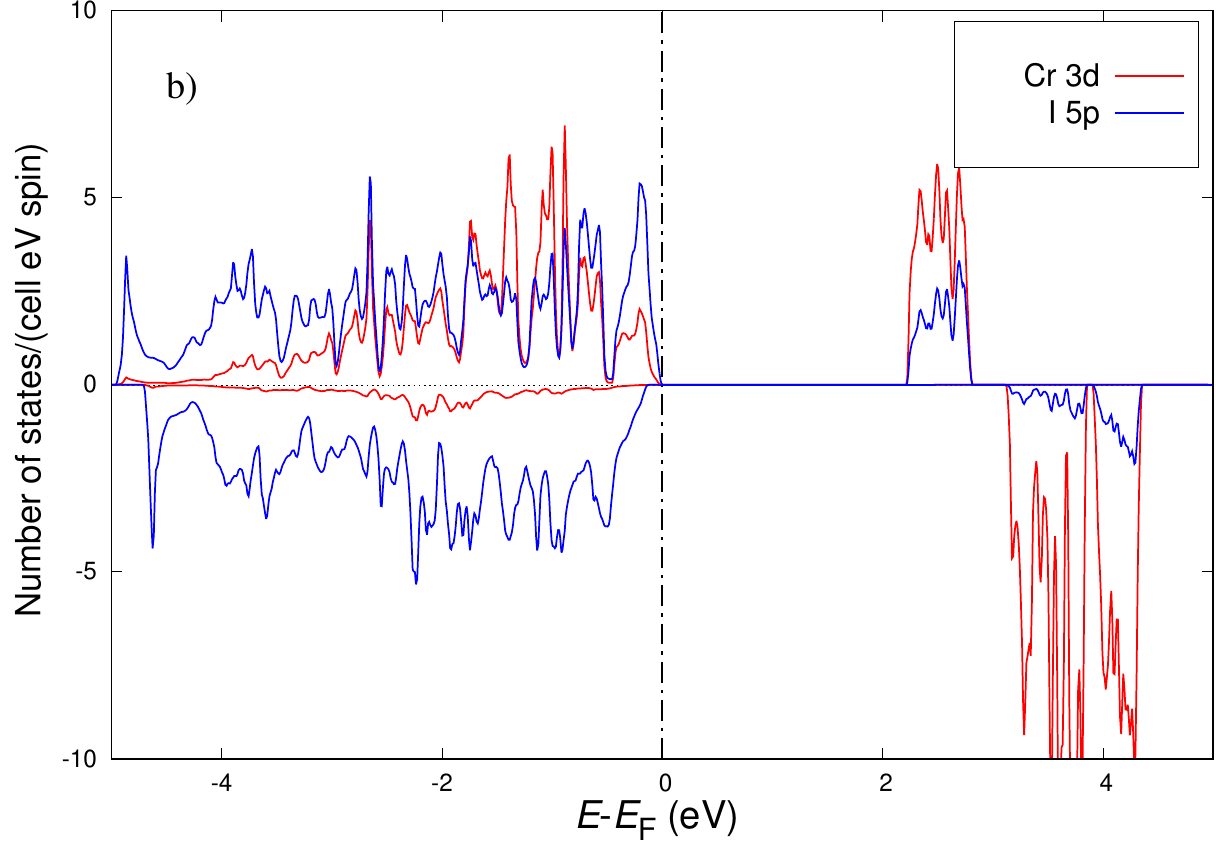} \\
   	  \includegraphics[width=.98\linewidth,clip]{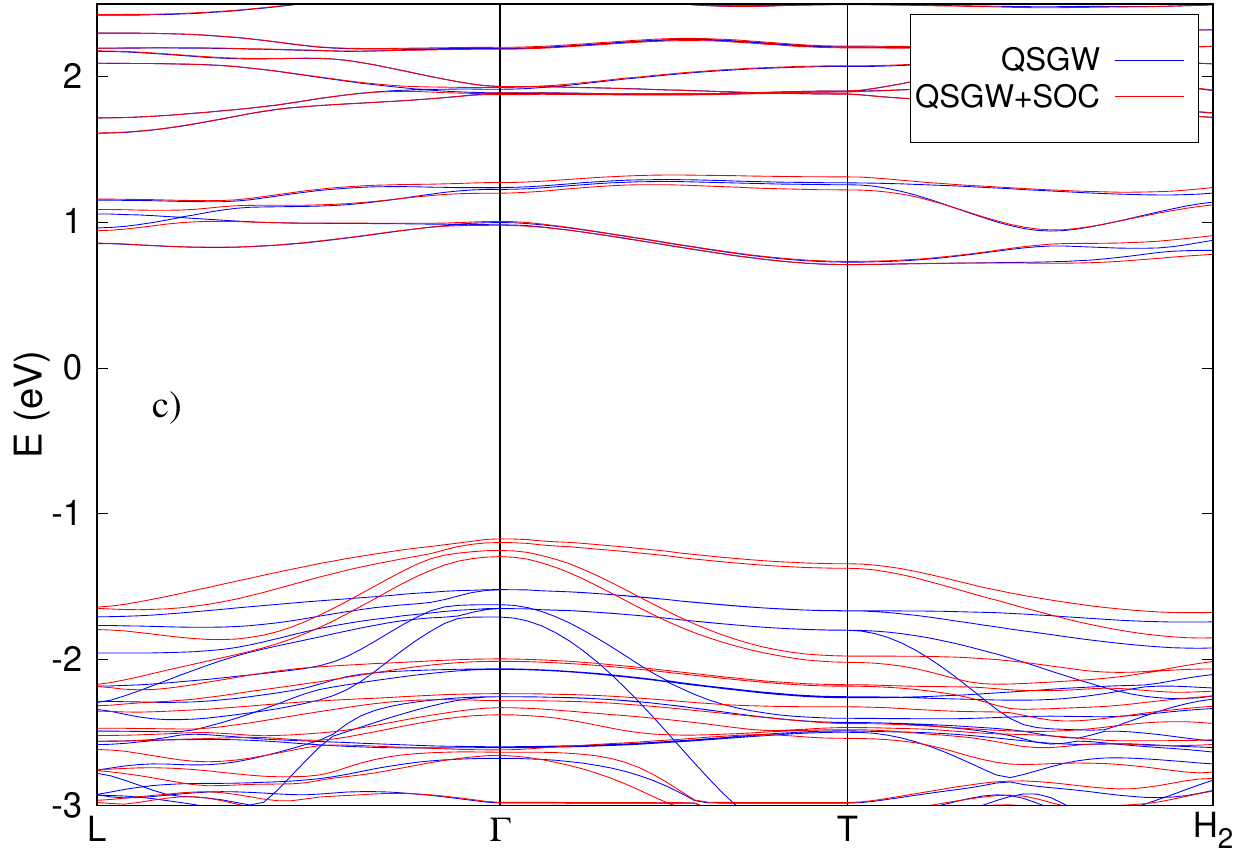}
	\end{tabular}%
	\caption{The partial density of states projected on Cr-$3d$ and I-$5p$ states in CrI$_{3}$ calculated within (a) DFT  and (b) $\qsgw$. (c) $\qsgw$ band structures of $\cri$ calculated with (red) and without (blue) SOC. }
	\label{fig:band_cri3}
\end{figure}

\paragraph*{$\cri$.---}

Figures~\ref{fig:band_cri3}(a) and \ref{fig:band_cri3}(b) show the PDOS of $\cri$ calculated in DFT and $\qsgw$, respectively, without SOC.
$\qsgw$ shifts up the unoccupied states in both spin channels, resulting in a larger $\eg$ than the one we obtain in DFT.
In the majority spin, the valence cation-$3d$ states are pushed down relative to the anion-$5p$ states, and the top of valence bands at $\Gamma$ becomes more dominated by anion-$p$ states.

\rFig{fig:band_cri3}(c) compares the $\qsgw$ band structures of $\cri$ calculated with and without SOC.
It clearly shows that SOC pushes up top valence bands around the $\Gamma$ point, resulting in a smaller $\eg$.
Within $\qsgw$, the top of majority-spin valence bands become more pure anion-$p$-like after $3d$ states are pushed down.
As a result, SOC has a stronger effect on decreasing $\eg$ in $\qsgw$ than in DFT.
Similar SOC effects are also found in $\vvi$ and $\cgt$.

\begin{figure}[bth]
\centering
\begin{tabular}{c}
\includegraphics[width=0.98\linewidth,clip]{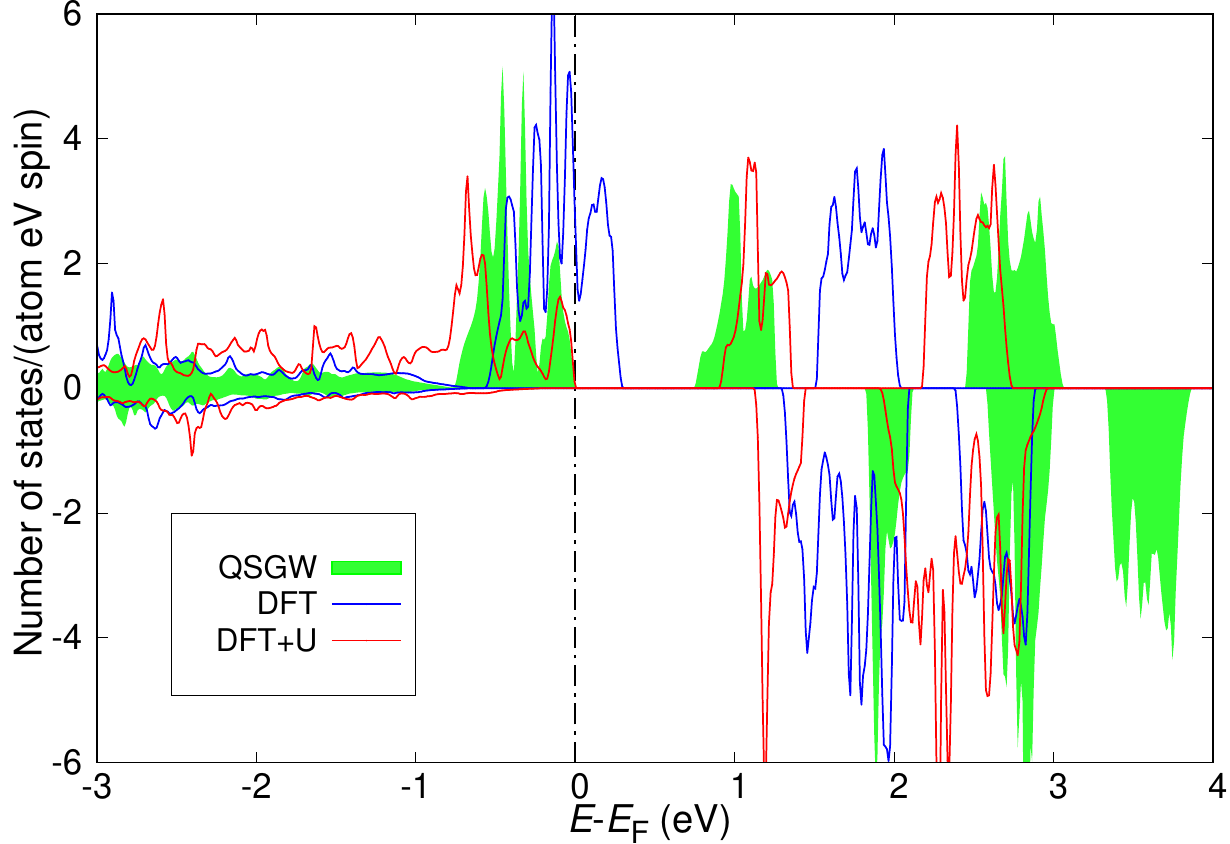}  
\end{tabular}%
\caption{The partial density of states projected on the V-$3d$ states in $\vvi$ within DFT (green shaded), DFT+$U$, and $\qsgw$.
DFT$+U$ calculation is performed using the AMF scheme. $U=\SI{2.7}{eV}$ is used so that the majority-spin V-$3d$ states peak at similar positions as in $\qsgw$. SOC is not included.} 
\label{fig:dos_vi3_cri3}
\end{figure}

\paragraph*{$\vvi$.---}

$\qsgw$ predicts that $\vvi$ is a semiconductor  while DFT incorrectly predicts it as a half metal.
$\eg$ obtained in $\qsgw$ is within the range of experimental values.
\rFig{fig:dos_vi3_cri3} shows the PDOS of $\vvi$ calculated within DFT, DFT$+U$, and $\qsgw$.
$\vvi$ has one less electron than $\cri$ in the formula unit.
Within DFT, the Fermi level intersects the majority-spin $t_{2g}$ states, resulting in a metallic state.
The $t_{2g}$ states consist of five roughly equally occupied $3d$ orbitals.
In contrast, remarkably, $\qsgw$ splits the $d_{z^2}$ states out of the occupied $t_{2g}$ states and shifts them above $\ef$.
Correspondingly, the remaining $t_{2g}$ states become more occupied, and a bandgap forms between the $d_{z^2}$ states and the other $t_{2g}$ states in the majority spin.
Other unoccupied $3d$ states also shift upward for both spins within $\qsgw$.

By adjusting $U$, DFT$+U$ can reproduce $\qsgw$ $\eg$ in $\vvi$.
However, as shown in \rfig{fig:dos_vi3_cri3}, a $U=\SI{2.7}{eV}$  may give similar positions of V-$3d$ DOS as in $\qsgw$ in the majority spin, but not in the minority one.
Moreover, the shapes of occupied majority-spin DOS change significantly in DFT$+U$, comparing those in $\qsgw$ and in DFT.

\paragraph*{DFT+$U$.---}
\begin{figure}[htb]
\centering
\begin{tabular}{c}
  \includegraphics[width=.98\linewidth,clip]{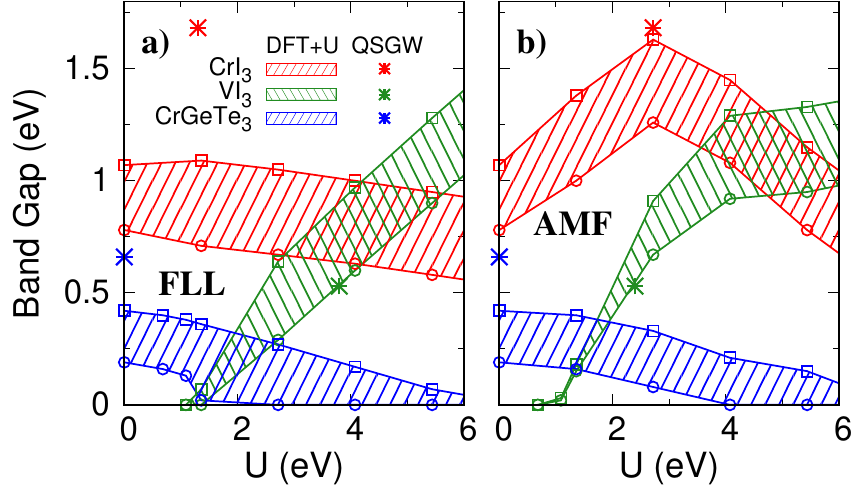} 
\end{tabular}%
\caption{ $\eg$ as a function of $U$ in $\cri$, $\vvi$, and $\cgt$, calculated  using the (a) fully-localized-limit scheme (FLL) and (b) around-the-mean-field (AMF) scheme.
The lower bound (open circles) and upper bound (open squares) of the shaded areas correspond to calculations with and without SOC, respectively.
$\qsgw$+SOC results are included to compare.
}
\label{fig:gap_vs_u}
\end{figure}

\rFig{fig:gap_vs_u} shows $\eg$ values calculated using two DFT+$U$ schemes, FLL and AMF, as a function of $U$, with and without SOC.
FLL and AMF give different $U$ dependences of $\eg$.
Within  FLL, $\eg$ values of $\cri$ and $\cgt$ decrease with increasing $U$, deviating further from experiments.
In $\vvi$, DFT$+U$ is not able to produce the experimental semiconducting state, especially with SOC, unless a sufficiently large $U$ is applied, e.g., \SIrange{2}{3}{eV} in AMF and \SIrange{3}{4}{eV} in FLL, respectively. 
Within AMF, $\eg$ values reach the maximum values with $U=2.7$ and \SI{6.8}{eV} in $\cri$ and $\vvi$, respectively, and then decrease.
In contrast to $\vvi$ and $\cri$, $\eg$ of $\cgt$ decreases with the increasing of $U$ value in both schemes.
Hence, DFT+$U$ is unable to increase $\eg$ in $\cgt$.

\begin{figure}[htb]
\centering
\begin{tabular}{c}
  \includegraphics[width=.98\linewidth,clip]{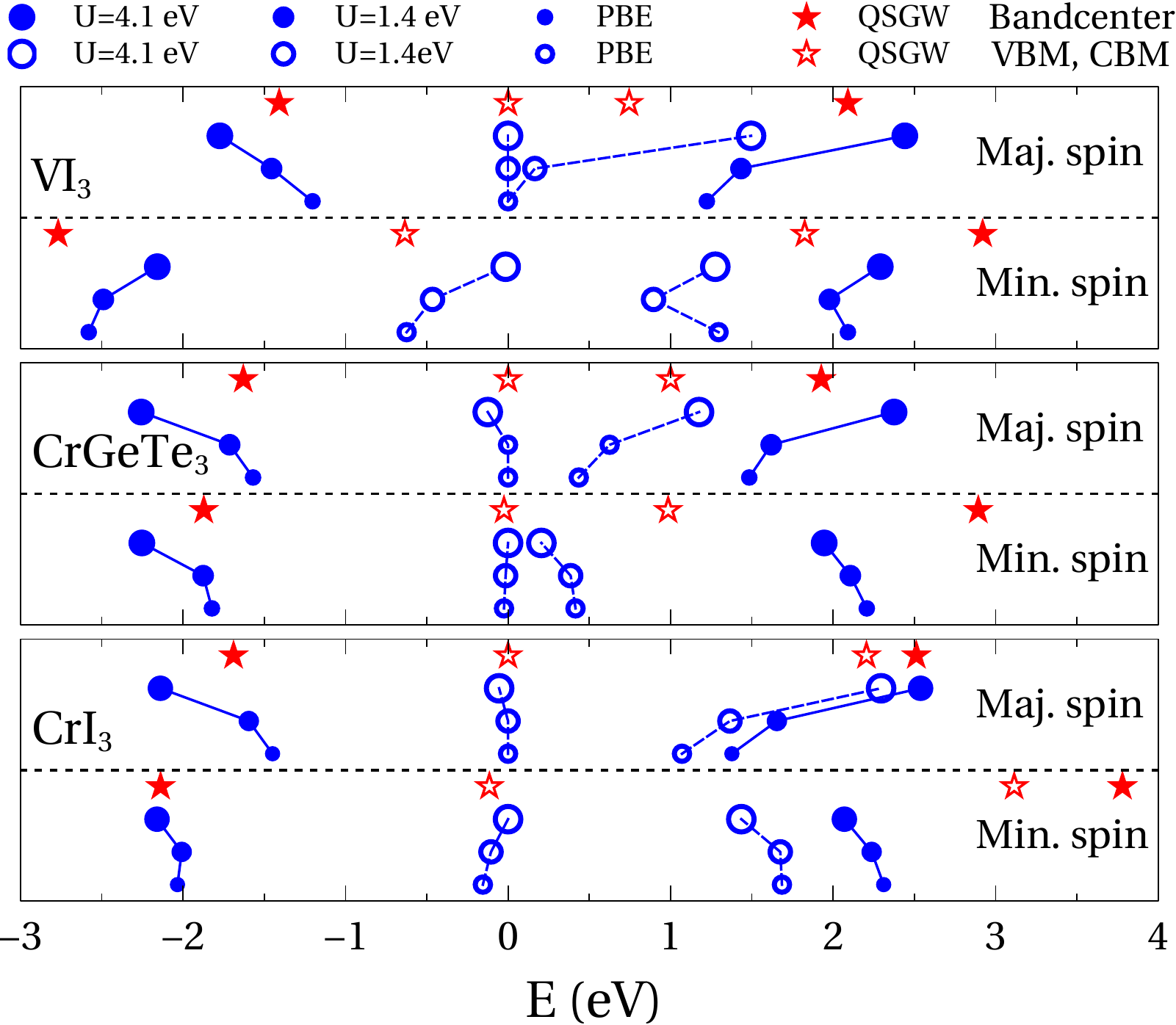}     
\end{tabular}%
\caption{ The CBM, VBM, and centers of $3d$ states in m2Dv in both spin channels calculated in DFT+$U$ and $\qsgw$.
  Band centers are denoted by solid circles (DFT+$U$) or stars ($\qsgw$) while CBM and VBM are denoted by open circles (DFT+$U$) or stars ($\qsgw$).
  The small, medium, and large circles represent $U=$\SIlist{0;1.4;4.1}{\eV}, respectively.
  The AMF scheme is used for DFT+$U$ calculation. SOC is not included.}
\label{fig:bandcenter_vs_u}
\end{figure}

To understand the behavior of $\eg$ dependence on $U$, we examine how electronic structures evolve with the increasing of $U$ in AMF.
\rFig{fig:bandcenter_vs_u} shows the $U$ dependence of the valence band maximum (VBM), the conduction band minimum (CBM), and the band centers of valence and conduction cation-$3d$ states in both spin channels, comparing with $\qsgw$ results.
For all three compounds, applying $U$ increases the gap and the distance between the centers of occupied and unoccupied $3d$ bands in the majority-spin but not 
the minority-spin channel.
This is clearly shown in ~\rfig{fig:dos_vi3_cri3} for the case of  $\vvi$ with $U=\SI{2.7}{\eV}$.
In $\cri$ and $\cgt$, a large $U$ pushes up the unoccupied $3d$ bands in the majority spin but lowers them in the minority spin.
When $U$ is sufficiently large, the unoccupied $3d$ states in the minority spin are shifted below those in the majority spin, and $\eg$ is determined by the exchange splitting instead of crystal-field splitting.
A similar trend is also observed in $\vvi$, but it occurs at a much larger $U$.
DOS calculated within DFT, DFT+$U$, and QSGW can be found in the Supplemental Material~\cite{sm}.

\emph{Can we mimic $\qsgw$ DOS by applying $U$ on cation-$d$ orbitals?}
Now we compare DFT+$U$ with $\qsgw$ DOS. 
As shown in~\rfig{fig:bandcenter_vs_u}, $\qsgw$ separates further, in comparison to DFT, the occupied and unoccupied states in both spin channels,
while DFT+$U$ only separates them in one spin channel.
Hence, within DFT$+U$, a single $U$ parameter is not able to mimic the $\qsgw$ $3d$ band centers simultaneously in both spin channels.
We also apply different $J$ values in DFT+$U$+$J$ calculations and are not able to reproduce $\qsgw$ $3d$ states in a satisfactory fashion as well. (Results of $\vvi$ are shown in Fig.~S10 in the Supplemental Material.)
Furthermore, VBM and CBM are the bonding and anti-bonding states made of cation-$3d$ and anion-$5p$ orbitals.
The positions of unoccupied cation-$3d$ bands relative to anion-$5p$ bands at VBM are not directly controlled by on-site $U$, which adjusts only occupied $3d$ bands, but by the off-site nonlocal potential that was naturally included in $\vxc_{\rm QSGW}$ within $\qsgw$.
Thus, there is no way that the DFT+$U$ can be used to mimic $\qsgw$ for these systems. 
It would be interesting to see whether extended Hubbard models, such as DFT+$U$+$V$~\cite{ricca2020arxiv}, can satisfactorily approximate such off-site correlations, especially with parameters determined systematically and automatically.

Although $\qsgw$ and DFT give the same or similar magnetic moments for all m2Dv we studied here, we expect different exchange couplings calculated in two methods, considering $\qsgw$'s profound effects on electronic structures. 
The anion-$5p$ weights at the top of valence bands are very different within two methods, suggesting that the corresponding superexchange couplings should differ as well.

Finally, $\qsgw$ is much more computational demanding in comparison with DFT.
Its efficiency needs to be improved for application to large-scale high-throughput calculations.
Recently, progress has been made in this direction.
For example, self-consistent $GW$ calculations using large unit cells with more than 50 atoms has become feasible~\cite{kutepov2019arxiv}.
By optimizing algorithms for the polarizability and the self-energy, Kutepov~\cite{kutepov2019arxiv} has shown the scaling of computational time is between linear and quadratic with respect to the system size, demonstrating the promising potential of its application on high-throughput computations.

\paragraph*{Conclusions.---}
We investigated the effects of the nonlocal exchange-correlation on the electronic structures of magnetic 2D van der Waals materials using the $\qsgw$ method.
$\qsgw$ correctly predicts the semiconducting states of $\vvi$ while DFT and $G_0W_0$ fail.
The corresponding calculated values are within the range of reported experimental values for $\cgt$ and $\vvi$, but larger than the experimental $\eg$ for $\cri$.
We also demonstrated that the simplistic DFT$+U$ method could not mimic the effects introduced by $\qsgw$, suggesting the importance of a more elaborate treatment of electron correlations in these systems.
Furthermore, considering the limitation of the DFT+$U$ method, the parameter-free and more universal $\qsgw$ method 
is more suitable to work as an engine in MI, providing a good independent-particle picture for high-throughput computations to search for new m2Dv.

\section*{Acknowledgments}
This work was supported by the U.S.~Department of Energy, Office of Science, Office of Basic Energy Sciences, Materials Sciences and Engineering Division, and Early Career Research Program.
Ames Laboratory is operated for the U.S.~Department of Energy by Iowa State University under Contract No.~DE-AC02-07CH11358.
This research used resources of the National Energy Research Scientific Computing Center (NERSC), a U.S.
Department of Energy Office of Science User Facility operated under Contract No.~DE-AC02-05CH11231.
T.~K. acknowledges the support from JSPS KAKENHI through Grant No.~17K05499, and the computing time provided by Research Institute for Information Technology (Kyushu University).

\appendix

\bibliography{aaa}

\begin{thebibliography}{54}%
\makeatletter
\providecommand \@ifxundefined [1]{%
 \@ifx{#1\undefined}
}%
\providecommand \@ifnum [1]{%
 \ifnum #1\expandafter \@firstoftwo
 \else \expandafter \@secondoftwo
 \fi
}%
\providecommand \@ifx [1]{%
 \ifx #1\expandafter \@firstoftwo
 \else \expandafter \@secondoftwo
 \fi
}%
\providecommand \natexlab [1]{#1}%
\providecommand \enquote  [1]{``#1''}%
\providecommand \bibnamefont  [1]{#1}%
\providecommand \bibfnamefont [1]{#1}%
\providecommand \citenamefont [1]{#1}%
\providecommand \href@noop [0]{\@secondoftwo}%
\providecommand \href [0]{\begingroup \@sanitize@url \@href}%
\providecommand \@href[1]{\@@startlink{#1}\@@href}%
\providecommand \@@href[1]{\endgroup#1\@@endlink}%
\providecommand \@sanitize@url [0]{\catcode `\\12\catcode `\$12\catcode
  `\&12\catcode `\#12\catcode `\^12\catcode `\_12\catcode `\%12\relax}%
\providecommand \@@startlink[1]{}%
\providecommand \@@endlink[0]{}%
\providecommand \url  [0]{\begingroup\@sanitize@url \@url }%
\providecommand \@url [1]{\endgroup\@href {#1}{\urlprefix }}%
\providecommand \urlprefix  [0]{URL }%
\providecommand \Eprint [0]{\href }%
\providecommand \doibase [0]{https://doi.org/}%
\providecommand \selectlanguage [0]{\@gobble}%
\providecommand \bibinfo  [0]{\@secondoftwo}%
\providecommand \bibfield  [0]{\@secondoftwo}%
\providecommand \translation [1]{[#1]}%
\providecommand \BibitemOpen [0]{}%
\providecommand \bibitemStop [0]{}%
\providecommand \bibitemNoStop [0]{.\EOS\space}%
\providecommand \EOS [0]{\spacefactor3000\relax}%
\providecommand \BibitemShut  [1]{\csname bibitem#1\endcsname}%
\let\auto@bib@innerbib\@empty
\bibitem [{\citenamefont {Zhong}\ \emph {et~al.}(2017)\citenamefont {Zhong},
  \citenamefont {Seyler}, \citenamefont {Linpeng}, \citenamefont {Cheng},
  \citenamefont {Sivadas}, \citenamefont {Huang}, \citenamefont {Schmidgall},
  \citenamefont {Taniguchi}, \citenamefont {Watanabe}, \citenamefont {McGuire},
  \citenamefont {Yao}, \citenamefont {Xiao}, \citenamefont {Fu},\ and\
  \citenamefont {Xu}}]{zhong2017sa}%
  \BibitemOpen
  \bibfield  {author} {\bibinfo {author} {\bibfnamefont {D.}~\bibnamefont
  {Zhong}}, \bibinfo {author} {\bibfnamefont {K.~L.}\ \bibnamefont {Seyler}},
  \bibinfo {author} {\bibfnamefont {X.}~\bibnamefont {Linpeng}}, \bibinfo
  {author} {\bibfnamefont {R.}~\bibnamefont {Cheng}}, \bibinfo {author}
  {\bibfnamefont {N.}~\bibnamefont {Sivadas}}, \bibinfo {author} {\bibfnamefont
  {B.}~\bibnamefont {Huang}}, \bibinfo {author} {\bibfnamefont
  {E.}~\bibnamefont {Schmidgall}}, \bibinfo {author} {\bibfnamefont
  {T.}~\bibnamefont {Taniguchi}}, \bibinfo {author} {\bibfnamefont
  {K.}~\bibnamefont {Watanabe}}, \bibinfo {author} {\bibfnamefont {M.~A.}\
  \bibnamefont {McGuire}}, \bibinfo {author} {\bibfnamefont {W.}~\bibnamefont
  {Yao}}, \bibinfo {author} {\bibfnamefont {D.}~\bibnamefont {Xiao}}, \bibinfo
  {author} {\bibfnamefont {K.-M.~C.}\ \bibnamefont {Fu}},\ and\ \bibinfo
  {author} {\bibfnamefont {X.}~\bibnamefont {Xu}},\ }\bibfield  {title}
  {\bibinfo {title} {{Van der Waals engineering of ferromagnetic semiconductor
  heterostructures for spin and valleytronics}},\ }\href@noop {} {\bibfield
  {journal} {\bibinfo  {journal} {Science Advances}\ }\textbf {\bibinfo
  {volume} {3}},\ \bibinfo {pages} {e1603113} (\bibinfo {year}
  {2017})}\BibitemShut {NoStop}%
\bibitem [{\citenamefont {Gong}\ \emph {et~al.}(2017)\citenamefont {Gong},
  \citenamefont {Li}, \citenamefont {Li}, \citenamefont {Ji}, \citenamefont
  {Stern}, \citenamefont {Xia}, \citenamefont {Cao}, \citenamefont {Bao},
  \citenamefont {Wang}, \citenamefont {Wang}, \citenamefont {Qiu},
  \citenamefont {Cava}, \citenamefont {Louie}, \citenamefont {Xia},\ and\
  \citenamefont {Zhang}}]{gong2017n}%
  \BibitemOpen
  \bibfield  {author} {\bibinfo {author} {\bibfnamefont {C.}~\bibnamefont
  {Gong}}, \bibinfo {author} {\bibfnamefont {L.}~\bibnamefont {Li}}, \bibinfo
  {author} {\bibfnamefont {Z.}~\bibnamefont {Li}}, \bibinfo {author}
  {\bibfnamefont {H.}~\bibnamefont {Ji}}, \bibinfo {author} {\bibfnamefont
  {A.}~\bibnamefont {Stern}}, \bibinfo {author} {\bibfnamefont
  {Y.}~\bibnamefont {Xia}}, \bibinfo {author} {\bibfnamefont {T.}~\bibnamefont
  {Cao}}, \bibinfo {author} {\bibfnamefont {W.}~\bibnamefont {Bao}}, \bibinfo
  {author} {\bibfnamefont {C.}~\bibnamefont {Wang}}, \bibinfo {author}
  {\bibfnamefont {Y.}~\bibnamefont {Wang}}, \bibinfo {author} {\bibfnamefont
  {Z.~Q.}\ \bibnamefont {Qiu}}, \bibinfo {author} {\bibfnamefont {R.~J.}\
  \bibnamefont {Cava}}, \bibinfo {author} {\bibfnamefont {S.~G.}\ \bibnamefont
  {Louie}}, \bibinfo {author} {\bibfnamefont {J.}~\bibnamefont {Xia}},\ and\
  \bibinfo {author} {\bibfnamefont {X.}~\bibnamefont {Zhang}},\ }\bibfield
  {title} {\bibinfo {title} {{Discovery of intrinsic ferromagnetism in
  two-dimensional van der Waals crystals}},\ }\href@noop {} {\bibfield
  {journal} {\bibinfo  {journal} {Nature}\ }\textbf {\bibinfo {volume} {546}},\
  \bibinfo {pages} {265} (\bibinfo {year} {2017})}\BibitemShut {NoStop}%
\bibitem [{\citenamefont {Huang}\ \emph {et~al.}(2017)\citenamefont {Huang},
  \citenamefont {Clark}, \citenamefont {Navarro-Moratalla}, \citenamefont
  {Klein}, \citenamefont {Cheng}, \citenamefont {Seyler}, \citenamefont
  {Zhong}, \citenamefont {Schmidgall}, \citenamefont {McGuire}, \citenamefont
  {Cobden}, \citenamefont {Yao}, \citenamefont {Xiao}, \citenamefont
  {Jarillo-Herrero},\ and\ \citenamefont {Xu}}]{huang2017n}%
  \BibitemOpen
  \bibfield  {author} {\bibinfo {author} {\bibfnamefont {B.}~\bibnamefont
  {Huang}}, \bibinfo {author} {\bibfnamefont {G.}~\bibnamefont {Clark}},
  \bibinfo {author} {\bibfnamefont {E.}~\bibnamefont {Navarro-Moratalla}},
  \bibinfo {author} {\bibfnamefont {D.~R.}\ \bibnamefont {Klein}}, \bibinfo
  {author} {\bibfnamefont {R.}~\bibnamefont {Cheng}}, \bibinfo {author}
  {\bibfnamefont {K.~L.}\ \bibnamefont {Seyler}}, \bibinfo {author}
  {\bibfnamefont {D.}~\bibnamefont {Zhong}}, \bibinfo {author} {\bibfnamefont
  {E.}~\bibnamefont {Schmidgall}}, \bibinfo {author} {\bibfnamefont {M.~A.}\
  \bibnamefont {McGuire}}, \bibinfo {author} {\bibfnamefont {D.~H.}\
  \bibnamefont {Cobden}}, \bibinfo {author} {\bibfnamefont {W.}~\bibnamefont
  {Yao}}, \bibinfo {author} {\bibfnamefont {D.}~\bibnamefont {Xiao}}, \bibinfo
  {author} {\bibfnamefont {P.}~\bibnamefont {Jarillo-Herrero}},\ and\ \bibinfo
  {author} {\bibfnamefont {X.}~\bibnamefont {Xu}},\ }\bibfield  {title}
  {\bibinfo {title} {{Layer-dependent ferromagnetism in a van der Waals crystal
  down to the monolayer limit}},\ }\href@noop {} {\bibfield  {journal}
  {\bibinfo  {journal} {Nature}\ }\textbf {\bibinfo {volume} {546}},\ \bibinfo
  {pages} {270} (\bibinfo {year} {2017})},\ \bibinfo {note}
  {letter}\BibitemShut {NoStop}%
\bibitem [{\citenamefont {Deng}\ \emph {et~al.}(2018)\citenamefont {Deng},
  \citenamefont {Yu}, \citenamefont {Song}, \citenamefont {Zhang},
  \citenamefont {Wang}, \citenamefont {Sun}, \citenamefont {Yi}, \citenamefont
  {Wu}, \citenamefont {Wu}, \citenamefont {Zhu}, \citenamefont {Wang},
  \citenamefont {Chen},\ and\ \citenamefont {Zhang}}]{deng2018n}%
  \BibitemOpen
  \bibfield  {author} {\bibinfo {author} {\bibfnamefont {Y.}~\bibnamefont
  {Deng}}, \bibinfo {author} {\bibfnamefont {Y.}~\bibnamefont {Yu}}, \bibinfo
  {author} {\bibfnamefont {Y.}~\bibnamefont {Song}}, \bibinfo {author}
  {\bibfnamefont {J.}~\bibnamefont {Zhang}}, \bibinfo {author} {\bibfnamefont
  {N.~Z.}\ \bibnamefont {Wang}}, \bibinfo {author} {\bibfnamefont
  {Z.}~\bibnamefont {Sun}}, \bibinfo {author} {\bibfnamefont {Y.}~\bibnamefont
  {Yi}}, \bibinfo {author} {\bibfnamefont {Y.~Z.}\ \bibnamefont {Wu}}, \bibinfo
  {author} {\bibfnamefont {S.}~\bibnamefont {Wu}}, \bibinfo {author}
  {\bibfnamefont {J.}~\bibnamefont {Zhu}}, \bibinfo {author} {\bibfnamefont
  {J.}~\bibnamefont {Wang}}, \bibinfo {author} {\bibfnamefont {X.~H.}\
  \bibnamefont {Chen}},\ and\ \bibinfo {author} {\bibfnamefont
  {Y.}~\bibnamefont {Zhang}},\ }\bibfield  {title} {\bibinfo {title}
  {{Gate-tunable room-temperature ferromagnetism in two-dimensional
  Fe$_3$GeTe$_2$}},\ }\href {https://doi.org/10.1038/s41586-018-0626-9}
  {\bibfield  {journal} {\bibinfo  {journal} {Nature}\ }\textbf {\bibinfo
  {volume} {563}},\ \bibinfo {pages} {94} (\bibinfo {year} {2018})}\BibitemShut
  {NoStop}%
\bibitem [{\citenamefont {Tian}\ \emph {et~al.}(2019)\citenamefont {Tian},
  \citenamefont {Zhang}, \citenamefont {Li}, \citenamefont {Ying},
  \citenamefont {Li}, \citenamefont {Zhang}, \citenamefont {Liu},\ and\
  \citenamefont {Lei}}]{tian2019jacs}%
  \BibitemOpen
  \bibfield  {author} {\bibinfo {author} {\bibfnamefont {S.}~\bibnamefont
  {Tian}}, \bibinfo {author} {\bibfnamefont {J.-F.}\ \bibnamefont {Zhang}},
  \bibinfo {author} {\bibfnamefont {C.}~\bibnamefont {Li}}, \bibinfo {author}
  {\bibfnamefont {T.}~\bibnamefont {Ying}}, \bibinfo {author} {\bibfnamefont
  {S.}~\bibnamefont {Li}}, \bibinfo {author} {\bibfnamefont {X.}~\bibnamefont
  {Zhang}}, \bibinfo {author} {\bibfnamefont {K.}~\bibnamefont {Liu}},\ and\
  \bibinfo {author} {\bibfnamefont {H.}~\bibnamefont {Lei}},\ }\bibfield
  {title} {\bibinfo {title} {{Ferromagnetic van der Waals Crystal VI$_3$}},\
  }\href {https://doi.org/10.1021/jacs.8b13584} {\bibfield  {journal} {\bibinfo
   {journal} {Journal of the American Chemical Society}\ }\textbf {\bibinfo
  {volume} {141}},\ \bibinfo {pages} {5326} (\bibinfo {year}
  {2019})}\BibitemShut {NoStop}%
\bibitem [{\citenamefont {He}\ \emph {et~al.}(2016)\citenamefont {He},
  \citenamefont {Ma}, \citenamefont {Lyu},\ and\ \citenamefont
  {Nachtigall}}]{he2016jmcc}%
  \BibitemOpen
  \bibfield  {author} {\bibinfo {author} {\bibfnamefont {J.}~\bibnamefont
  {He}}, \bibinfo {author} {\bibfnamefont {S.}~\bibnamefont {Ma}}, \bibinfo
  {author} {\bibfnamefont {P.}~\bibnamefont {Lyu}},\ and\ \bibinfo {author}
  {\bibfnamefont {P.}~\bibnamefont {Nachtigall}},\ }\bibfield  {title}
  {\bibinfo {title} {{Unusual Dirac half-metallicity with intrinsic
  ferromagnetism in vanadium trihalide monolayers}},\ }\href
  {https://doi.org/10.1039/C6TC00409A} {\bibfield  {journal} {\bibinfo
  {journal} {J. Mater. Chem. C}\ }\textbf {\bibinfo {volume} {4}},\ \bibinfo
  {pages} {2518} (\bibinfo {year} {2016})}\BibitemShut {NoStop}%
\bibitem [{\citenamefont {Kong}\ \emph {et~al.}(2019)\citenamefont {Kong},
  \citenamefont {Stolze}, \citenamefont {Timmons}, \citenamefont {Tao},
  \citenamefont {Ni}, \citenamefont {Guo}, \citenamefont {Yang}, \citenamefont
  {Prozorov},\ and\ \citenamefont {Cava}}]{kong2019am}%
  \BibitemOpen
  \bibfield  {author} {\bibinfo {author} {\bibfnamefont {T.}~\bibnamefont
  {Kong}}, \bibinfo {author} {\bibfnamefont {K.}~\bibnamefont {Stolze}},
  \bibinfo {author} {\bibfnamefont {E.~I.}\ \bibnamefont {Timmons}}, \bibinfo
  {author} {\bibfnamefont {J.}~\bibnamefont {Tao}}, \bibinfo {author}
  {\bibfnamefont {D.}~\bibnamefont {Ni}}, \bibinfo {author} {\bibfnamefont
  {S.}~\bibnamefont {Guo}}, \bibinfo {author} {\bibfnamefont {Z.}~\bibnamefont
  {Yang}}, \bibinfo {author} {\bibfnamefont {R.}~\bibnamefont {Prozorov}},\
  and\ \bibinfo {author} {\bibfnamefont {R.~J.}\ \bibnamefont {Cava}},\
  }\bibfield  {title} {\bibinfo {title} {{VI$_3$---a New Layered Ferromagnetic
  Semiconductor}},\ }\href {https://doi.org/10.1002/adma.201808074} {\bibfield
  {journal} {\bibinfo  {journal} {Advanced Materials}\ }\textbf {\bibinfo
  {volume} {31}},\ \bibinfo {pages} {1808074} (\bibinfo {year}
  {2019})}\BibitemShut {NoStop}%
\bibitem [{\citenamefont {Baidya}\ \emph {et~al.}(2018)\citenamefont {Baidya},
  \citenamefont {Yu},\ and\ \citenamefont {Kim}}]{baidya2018prb}%
  \BibitemOpen
  \bibfield  {author} {\bibinfo {author} {\bibfnamefont {S.}~\bibnamefont
  {Baidya}}, \bibinfo {author} {\bibfnamefont {J.}~\bibnamefont {Yu}},\ and\
  \bibinfo {author} {\bibfnamefont {C.~H.}\ \bibnamefont {Kim}},\ }\bibfield
  {title} {\bibinfo {title} {{Tunable magnetic topological insulating phases in
  monolayer CrI$_{3}$}},\ }\href {https://doi.org/10.1103/PhysRevB.98.155148}
  {\bibfield  {journal} {\bibinfo  {journal} {Phys. Rev. B}\ }\textbf {\bibinfo
  {volume} {98}},\ \bibinfo {pages} {155148} (\bibinfo {year}
  {2018})}\BibitemShut {NoStop}%
\bibitem [{\citenamefont {Fang}\ \emph {et~al.}(2018)\citenamefont {Fang},
  \citenamefont {Wu}, \citenamefont {Zhu},\ and\ \citenamefont
  {Guo}}]{fang2018prb}%
  \BibitemOpen
  \bibfield  {author} {\bibinfo {author} {\bibfnamefont {Y.}~\bibnamefont
  {Fang}}, \bibinfo {author} {\bibfnamefont {S.}~\bibnamefont {Wu}}, \bibinfo
  {author} {\bibfnamefont {Z.-Z.}\ \bibnamefont {Zhu}},\ and\ \bibinfo {author}
  {\bibfnamefont {G.-Y.}\ \bibnamefont {Guo}},\ }\bibfield  {title} {\bibinfo
  {title} {{Large magneto-optical effects and magnetic anisotropy energy in
  two-dimensional Cr$_{2}$Ge$_{2}$Te$_{6}$}},\ }\href
  {https://doi.org/10.1103/PhysRevB.98.125416} {\bibfield  {journal} {\bibinfo
  {journal} {Phys. Rev. B}\ }\textbf {\bibinfo {volume} {98}},\ \bibinfo
  {pages} {125416} (\bibinfo {year} {2018})}\BibitemShut {NoStop}%
\bibitem [{\citenamefont {Jiang}\ \emph {et~al.}(2018)\citenamefont {Jiang},
  \citenamefont {Li}, \citenamefont {Liao}, \citenamefont {Zhao},\ and\
  \citenamefont {Zhong}}]{jiang2018nl}%
  \BibitemOpen
  \bibfield  {author} {\bibinfo {author} {\bibfnamefont {P.}~\bibnamefont
  {Jiang}}, \bibinfo {author} {\bibfnamefont {L.}~\bibnamefont {Li}}, \bibinfo
  {author} {\bibfnamefont {Z.}~\bibnamefont {Liao}}, \bibinfo {author}
  {\bibfnamefont {Y.~X.}\ \bibnamefont {Zhao}},\ and\ \bibinfo {author}
  {\bibfnamefont {Z.}~\bibnamefont {Zhong}},\ }\bibfield  {title} {\bibinfo
  {title} {{Spin Direction-Controlled Electronic Band Structure in
  Two-Dimensional Ferromagnetic CrI$_3$}},\ }\href
  {https://doi.org/10.1021/acs.nanolett.8b01125} {\bibfield  {journal}
  {\bibinfo  {journal} {Nano Letters}\ }\textbf {\bibinfo {volume} {18}},\
  \bibinfo {pages} {3844} (\bibinfo {year} {2018})},\ \bibinfo {note} {pMID:
  29783842}\BibitemShut {NoStop}%
\bibitem [{\citenamefont {Kulish}\ and\ \citenamefont
  {Huang}(2017)}]{kulish2017jmcc}%
  \BibitemOpen
  \bibfield  {author} {\bibinfo {author} {\bibfnamefont {V.~V.}\ \bibnamefont
  {Kulish}}\ and\ \bibinfo {author} {\bibfnamefont {W.}~\bibnamefont {Huang}},\
  }\bibfield  {title} {\bibinfo {title} {Single-layer metal halides {M$X_2$
  ($X$ = Cl, Br, I)}: stability and tunable magnetism from first principles and
  {Monte Carlo} simulations},\ }\href {https://doi.org/10.1039/C7TC02664A}
  {\bibfield  {journal} {\bibinfo  {journal} {J. Mater. Chem. C}\ }\textbf
  {\bibinfo {volume} {5}},\ \bibinfo {pages} {8734} (\bibinfo {year}
  {2017})}\BibitemShut {NoStop}%
\bibitem [{\citenamefont {Sivadas}\ \emph {et~al.}(2018)\citenamefont
  {Sivadas}, \citenamefont {Okamoto}, \citenamefont {Xu}, \citenamefont
  {Fennie},\ and\ \citenamefont {Xiao}}]{sivadas2018nl}%
  \BibitemOpen
  \bibfield  {author} {\bibinfo {author} {\bibfnamefont {N.}~\bibnamefont
  {Sivadas}}, \bibinfo {author} {\bibfnamefont {S.}~\bibnamefont {Okamoto}},
  \bibinfo {author} {\bibfnamefont {X.}~\bibnamefont {Xu}}, \bibinfo {author}
  {\bibfnamefont {C.~J.}\ \bibnamefont {Fennie}},\ and\ \bibinfo {author}
  {\bibfnamefont {D.}~\bibnamefont {Xiao}},\ }\bibfield  {title} {\bibinfo
  {title} {{Stacking-Dependent Magnetism in Bilayer CrI$_3$}},\ }\href
  {https://doi.org/10.1021/acs.nanolett.8b03321} {\bibfield  {journal}
  {\bibinfo  {journal} {Nano Letters}\ }\textbf {\bibinfo {volume} {18}},\
  \bibinfo {pages} {7658} (\bibinfo {year} {2018})}\BibitemShut {NoStop}%
\bibitem [{\citenamefont {Menichetti}\ \emph {et~al.}(2019)\citenamefont
  {Menichetti}, \citenamefont {Calandra},\ and\ \citenamefont
  {Polini}}]{menichetti20192m}%
  \BibitemOpen
  \bibfield  {author} {\bibinfo {author} {\bibfnamefont {G.}~\bibnamefont
  {Menichetti}}, \bibinfo {author} {\bibfnamefont {M.}~\bibnamefont
  {Calandra}},\ and\ \bibinfo {author} {\bibfnamefont {M.}~\bibnamefont
  {Polini}},\ }\bibfield  {title} {\bibinfo {title} {{Electronic structure and
  magnetic properties of few-layer Cr$_2$Ge$_2$Te$_6$: the key role of nonlocal
  electron{\textendash}electron interaction effects}},\ }\href
  {https://doi.org/10.1088/2053-1583/ab2f06} {\bibfield  {journal} {\bibinfo
  {journal} {2D Materials}\ }\textbf {\bibinfo {volume} {6}},\ \bibinfo {pages}
  {045042} (\bibinfo {year} {2019})}\BibitemShut {NoStop}%
\bibitem [{\citenamefont {Torelli}\ and\ \citenamefont
  {Olsen}(2018)}]{torelli20182m}%
  \BibitemOpen
  \bibfield  {author} {\bibinfo {author} {\bibfnamefont {D.}~\bibnamefont
  {Torelli}}\ and\ \bibinfo {author} {\bibfnamefont {T.}~\bibnamefont
  {Olsen}},\ }\bibfield  {title} {\bibinfo {title} {{Calculating critical
  temperatures for ferromagnetic order in two-dimensional materials}},\ }\href
  {https://doi.org/10.1088/2053-1583/aaf06d} {\bibfield  {journal} {\bibinfo
  {journal} {2D Materials}\ }\textbf {\bibinfo {volume} {6}},\ \bibinfo {pages}
  {015028} (\bibinfo {year} {2018})}\BibitemShut {NoStop}%
\bibitem [{\citenamefont {Lado}\ and\ \citenamefont
  {Fern{\'{a}}ndez-Rossier}(2017)}]{lado20172m}%
  \BibitemOpen
  \bibfield  {author} {\bibinfo {author} {\bibfnamefont {J.~L.}\ \bibnamefont
  {Lado}}\ and\ \bibinfo {author} {\bibfnamefont {J.}~\bibnamefont
  {Fern{\'{a}}ndez-Rossier}},\ }\bibfield  {title} {\bibinfo {title} {On the
  origin of magnetic anisotropy in two dimensional {CrI$_3$}},\ }\href
  {https://doi.org/10.1088/2053-1583/aa75ed} {\bibfield  {journal} {\bibinfo
  {journal} {2D Materials}\ }\textbf {\bibinfo {volume} {4}},\ \bibinfo {pages}
  {035002} (\bibinfo {year} {2017})}\BibitemShut {NoStop}%
\bibitem [{\citenamefont {Mounet}\ \emph {et~al.}(2018)\citenamefont {Mounet},
  \citenamefont {Gibertini}, \citenamefont {Schwaller}, \citenamefont {Campi},
  \citenamefont {Merkys}, \citenamefont {Marrazzo}, \citenamefont {Sohier},
  \citenamefont {Castelli}, \citenamefont {Cepellotti}, \citenamefont {Pizzi},\
  and\ \citenamefont {Marzari}}]{mounet2018nn}%
  \BibitemOpen
  \bibfield  {author} {\bibinfo {author} {\bibfnamefont {N.}~\bibnamefont
  {Mounet}}, \bibinfo {author} {\bibfnamefont {M.}~\bibnamefont {Gibertini}},
  \bibinfo {author} {\bibfnamefont {P.}~\bibnamefont {Schwaller}}, \bibinfo
  {author} {\bibfnamefont {D.}~\bibnamefont {Campi}}, \bibinfo {author}
  {\bibfnamefont {A.}~\bibnamefont {Merkys}}, \bibinfo {author} {\bibfnamefont
  {A.}~\bibnamefont {Marrazzo}}, \bibinfo {author} {\bibfnamefont
  {T.}~\bibnamefont {Sohier}}, \bibinfo {author} {\bibfnamefont {I.~E.}\
  \bibnamefont {Castelli}}, \bibinfo {author} {\bibfnamefont {A.}~\bibnamefont
  {Cepellotti}}, \bibinfo {author} {\bibfnamefont {G.}~\bibnamefont {Pizzi}},\
  and\ \bibinfo {author} {\bibfnamefont {N.}~\bibnamefont {Marzari}},\
  }\bibfield  {title} {\bibinfo {title} {{Two-dimensional materials from
  high-throughput computational exfoliation of experimentally known
  compounds}},\ }\href {https://doi.org/10.1038/s41565-017-0035-5} {\bibfield
  {journal} {\bibinfo  {journal} {Nature Nanotechnology}\ }\textbf {\bibinfo
  {volume} {13}},\ \bibinfo {pages} {246} (\bibinfo {year} {2018})}\BibitemShut
  {NoStop}%
\bibitem [{\citenamefont {Jang}\ \emph {et~al.}(2019)\citenamefont {Jang},
  \citenamefont {Jeong}, \citenamefont {Yoon}, \citenamefont {Ryee},\ and\
  \citenamefont {Han}}]{jang2019prm}%
  \BibitemOpen
  \bibfield  {author} {\bibinfo {author} {\bibfnamefont {S.~W.}\ \bibnamefont
  {Jang}}, \bibinfo {author} {\bibfnamefont {M.~Y.}\ \bibnamefont {Jeong}},
  \bibinfo {author} {\bibfnamefont {H.}~\bibnamefont {Yoon}}, \bibinfo {author}
  {\bibfnamefont {S.}~\bibnamefont {Ryee}},\ and\ \bibinfo {author}
  {\bibfnamefont {M.~J.}\ \bibnamefont {Han}},\ }\bibfield  {title} {\bibinfo
  {title} {{Microscopic understanding of magnetic interactions in bilayer
  CrI$_3$}},\ }\href {https://doi.org/10.1103/PhysRevMaterials.3.031001}
  {\bibfield  {journal} {\bibinfo  {journal} {Physical Review Materials}\
  }\textbf {\bibinfo {volume} {3}},\ \bibinfo {pages} {031001} (\bibinfo {year}
  {2019})}\BibitemShut {NoStop}%
\bibitem [{\citenamefont {Hao}\ \emph {et~al.}(2018)\citenamefont {Hao},
  \citenamefont {Li}, \citenamefont {Zhang}, \citenamefont {Li}, \citenamefont
  {Lin}, \citenamefont {Luo}, \citenamefont {Sun}, \citenamefont {Liu},\ and\
  \citenamefont {Wang}}]{hao2018sb}%
  \BibitemOpen
  \bibfield  {author} {\bibinfo {author} {\bibfnamefont {Z.}~\bibnamefont
  {Hao}}, \bibinfo {author} {\bibfnamefont {H.}~\bibnamefont {Li}}, \bibinfo
  {author} {\bibfnamefont {S.}~\bibnamefont {Zhang}}, \bibinfo {author}
  {\bibfnamefont {X.}~\bibnamefont {Li}}, \bibinfo {author} {\bibfnamefont
  {G.}~\bibnamefont {Lin}}, \bibinfo {author} {\bibfnamefont {X.}~\bibnamefont
  {Luo}}, \bibinfo {author} {\bibfnamefont {Y.}~\bibnamefont {Sun}}, \bibinfo
  {author} {\bibfnamefont {Z.}~\bibnamefont {Liu}},\ and\ \bibinfo {author}
  {\bibfnamefont {Y.}~\bibnamefont {Wang}},\ }\bibfield  {title} {\bibinfo
  {title} {{Atomic scale electronic structure of the ferromagnetic
  semiconductor Cr$_2$Ge$_2$Te$_6$}},\ }\href
  {https://doi.org/https://doi.org/10.1016/j.scib.2018.05.034} {\bibfield
  {journal} {\bibinfo  {journal} {Science Bulletin}\ }\textbf {\bibinfo
  {volume} {63}},\ \bibinfo {pages} {825 } (\bibinfo {year}
  {2018})}\BibitemShut {NoStop}%
\bibitem [{\citenamefont {Li}\ and\ \citenamefont {Yang}(2014)}]{li2014jmcc}%
  \BibitemOpen
  \bibfield  {author} {\bibinfo {author} {\bibfnamefont {X.}~\bibnamefont
  {Li}}\ and\ \bibinfo {author} {\bibfnamefont {J.}~\bibnamefont {Yang}},\
  }\bibfield  {title} {\bibinfo {title} {{Cr$X$Te$_3$ ($X$ = Si, Ge)}
  nanosheets: two dimensional intrinsic ferromagnetic semiconductors},\ }\href
  {https://doi.org/10.1039/C4TC01193G} {\bibfield  {journal} {\bibinfo
  {journal} {J. Mater. Chem. C}\ }\textbf {\bibinfo {volume} {2}},\ \bibinfo
  {pages} {7071} (\bibinfo {year} {2014})}\BibitemShut {NoStop}%
\bibitem [{\citenamefont {Son}\ \emph {et~al.}(2019)\citenamefont {Son},
  \citenamefont {Coak}, \citenamefont {Lee}, \citenamefont {Kim}, \citenamefont
  {Kim}, \citenamefont {Hamidov}, \citenamefont {Cho}, \citenamefont {Liu},
  \citenamefont {Jarvis}, \citenamefont {Brown}, \citenamefont {Kim},
  \citenamefont {Park}, \citenamefont {Khomskii}, \citenamefont {Saxena},\ and\
  \citenamefont {Park}}]{son2019prb}%
  \BibitemOpen
  \bibfield  {author} {\bibinfo {author} {\bibfnamefont {S.}~\bibnamefont
  {Son}}, \bibinfo {author} {\bibfnamefont {M.~J.}\ \bibnamefont {Coak}},
  \bibinfo {author} {\bibfnamefont {N.}~\bibnamefont {Lee}}, \bibinfo {author}
  {\bibfnamefont {J.}~\bibnamefont {Kim}}, \bibinfo {author} {\bibfnamefont
  {T.~Y.}\ \bibnamefont {Kim}}, \bibinfo {author} {\bibfnamefont
  {H.}~\bibnamefont {Hamidov}}, \bibinfo {author} {\bibfnamefont
  {H.}~\bibnamefont {Cho}}, \bibinfo {author} {\bibfnamefont {C.}~\bibnamefont
  {Liu}}, \bibinfo {author} {\bibfnamefont {D.~M.}\ \bibnamefont {Jarvis}},
  \bibinfo {author} {\bibfnamefont {P.~A.~C.}\ \bibnamefont {Brown}}, \bibinfo
  {author} {\bibfnamefont {J.~H.}\ \bibnamefont {Kim}}, \bibinfo {author}
  {\bibfnamefont {C.-H.}\ \bibnamefont {Park}}, \bibinfo {author}
  {\bibfnamefont {D.~I.}\ \bibnamefont {Khomskii}}, \bibinfo {author}
  {\bibfnamefont {S.~S.}\ \bibnamefont {Saxena}},\ and\ \bibinfo {author}
  {\bibfnamefont {J.-G.}\ \bibnamefont {Park}},\ }\bibfield  {title} {\bibinfo
  {title} {{Bulk properties of the van der Waals hard ferromagnet VI$_{3}$}},\
  }\href {https://doi.org/10.1103/PhysRevB.99.041402} {\bibfield  {journal}
  {\bibinfo  {journal} {Phys. Rev. B}\ }\textbf {\bibinfo {volume} {99}},\
  \bibinfo {pages} {041402} (\bibinfo {year} {2019})}\BibitemShut {NoStop}%
\bibitem [{\citenamefont {Deguchi}\ \emph {et~al.}(2016)\citenamefont
  {Deguchi}, \citenamefont {Sato}, \citenamefont {Kino},\ and\ \citenamefont
  {Kotani}}]{deguchi2016jjap}%
  \BibitemOpen
  \bibfield  {author} {\bibinfo {author} {\bibfnamefont {D.}~\bibnamefont
  {Deguchi}}, \bibinfo {author} {\bibfnamefont {K.}~\bibnamefont {Sato}},
  \bibinfo {author} {\bibfnamefont {H.}~\bibnamefont {Kino}},\ and\ \bibinfo
  {author} {\bibfnamefont {T.}~\bibnamefont {Kotani}},\ }\bibfield  {title}
  {\bibinfo {title} {{Accurate energy bands calculated by the hybrid
  quasiparticle self-consistent $GW$ method implemented in the \textsc{ecalj}
  package}},\ }\href {https://doi.org/10.7567/jjap.55.051201} {\bibfield
  {journal} {\bibinfo  {journal} {Japanese Journal of Applied Physics}\
  }\textbf {\bibinfo {volume} {55}},\ \bibinfo {pages} {051201} (\bibinfo
  {year} {2016})}\BibitemShut {NoStop}%
\bibitem [{\citenamefont {Kotani}\ \emph {et~al.}(2007)\citenamefont {Kotani},
  \citenamefont {van Schilfgaarde},\ and\ \citenamefont
  {Faleev}}]{kotani2007prb}%
  \BibitemOpen
  \bibfield  {author} {\bibinfo {author} {\bibfnamefont {T.}~\bibnamefont
  {Kotani}}, \bibinfo {author} {\bibfnamefont {M.}~\bibnamefont {van
  Schilfgaarde}},\ and\ \bibinfo {author} {\bibfnamefont {S.~V.}\ \bibnamefont
  {Faleev}},\ }\bibfield  {title} {\bibinfo {title} {Quasiparticle
  self-consistent {$GW$} method: {A} basis for the independent-particle
  approximation},\ }\href {https://doi.org/10.1103/PhysRevB.76.165106}
  {\bibfield  {journal} {\bibinfo  {journal} {Phys. Rev. B}\ }\textbf {\bibinfo
  {volume} {76}},\ \bibinfo {pages} {165106} (\bibinfo {year}
  {2007})}\BibitemShut {NoStop}%
\bibitem [{\citenamefont {Kotani}(2014)}]{kotani2014jpsj}%
  \BibitemOpen
  \bibfield  {author} {\bibinfo {author} {\bibfnamefont {T.}~\bibnamefont
  {Kotani}},\ }\bibfield  {title} {\bibinfo {title} {Quasiparticle
  {Self-Consistent $GW$ Method Based} on the {Augmented Plane-Wave and
  Muffin-Tin Orbital Method}},\ }\href@noop {} {\bibfield  {journal} {\bibinfo
  {journal} {Journal of the Physical Society of Japan}\ }\textbf {\bibinfo
  {volume} {83}},\ \bibinfo {pages} {094711} (\bibinfo {year}
  {2014})}\BibitemShut {NoStop}%
\bibitem [{\citenamefont {Hedin}\ and\ \citenamefont
  {Lundqvist}(1969)}]{hedin1969book}%
  \BibitemOpen
  \bibfield  {author} {\bibinfo {author} {\bibfnamefont {L.}~\bibnamefont
  {Hedin}}\ and\ \bibinfo {author} {\bibfnamefont {S.}~\bibnamefont
  {Lundqvist}},\ }\href@noop {} {\emph {\bibinfo {title} {{Effects of
  Electron-Electron and Electron-Phonon interactions on the One-Electron States
  of Solids}}}},\ Vol.~\bibinfo {volume} {12}\ (\bibinfo  {publisher} {Oxford
  university press New York},\ \bibinfo {year} {1969})\BibitemShut {NoStop}%
\bibitem [{\citenamefont {Aryasetiawan}\ and\ \citenamefont
  {Gunnarsson}(1998)}]{aryasetiawan1998ropi}%
  \BibitemOpen
  \bibfield  {author} {\bibinfo {author} {\bibfnamefont {F.}~\bibnamefont
  {Aryasetiawan}}\ and\ \bibinfo {author} {\bibfnamefont {O.}~\bibnamefont
  {Gunnarsson}},\ }\bibfield  {title} {\bibinfo {title} {{The $GW$ method}},\
  }\href {https://doi.org/10.1088/0034-4885/61/3/002} {\bibfield  {journal}
  {\bibinfo  {journal} {Reports on Progress in Physics}\ }\textbf {\bibinfo
  {volume} {61}},\ \bibinfo {pages} {237} (\bibinfo {year} {1998})}\BibitemShut
  {NoStop}%
\bibitem [{\citenamefont {Kutepov}(2016)}]{kutepov2016prb}%
  \BibitemOpen
  \bibfield  {author} {\bibinfo {author} {\bibfnamefont {A.~L.}\ \bibnamefont
  {Kutepov}},\ }\bibfield  {title} {\bibinfo {title} {{Electronic structure of
  Na, K, Si, and LiF from self-consistent solution of Hedin's equations
  including vertex corrections}},\ }\href
  {https://doi.org/10.1103/PhysRevB.94.155101} {\bibfield  {journal} {\bibinfo
  {journal} {Phys. Rev. B}\ }\textbf {\bibinfo {volume} {94}},\ \bibinfo
  {pages} {155101} (\bibinfo {year} {2016})}\BibitemShut {NoStop}%
\bibitem [{\citenamefont {Hybertsen}\ and\ \citenamefont
  {Louie}(1986)}]{hybertsen1986prb}%
  \BibitemOpen
  \bibfield  {author} {\bibinfo {author} {\bibfnamefont {M.~S.}\ \bibnamefont
  {Hybertsen}}\ and\ \bibinfo {author} {\bibfnamefont {S.~G.}\ \bibnamefont
  {Louie}},\ }\bibfield  {title} {\bibinfo {title} {{Electron correlation in
  semiconductors and insulators: Band gaps and quasiparticle energies}},\
  }\href {https://doi.org/10.1103/PhysRevB.34.5390} {\bibfield  {journal}
  {\bibinfo  {journal} {Phys. Rev. B}\ }\textbf {\bibinfo {volume} {34}},\
  \bibinfo {pages} {5390} (\bibinfo {year} {1986})}\BibitemShut {NoStop}%
\bibitem [{\citenamefont {Shishkin}\ \emph {et~al.}(2007)\citenamefont
  {Shishkin}, \citenamefont {Marsman},\ and\ \citenamefont
  {Kresse}}]{shishkin2007prl}%
  \BibitemOpen
  \bibfield  {author} {\bibinfo {author} {\bibfnamefont {M.}~\bibnamefont
  {Shishkin}}, \bibinfo {author} {\bibfnamefont {M.}~\bibnamefont {Marsman}},\
  and\ \bibinfo {author} {\bibfnamefont {G.}~\bibnamefont {Kresse}},\
  }\bibfield  {title} {\bibinfo {title} {{Accurate Quasiparticle Spectra from
  Self-Consistent $GW$ Calculations with Vertex Corrections}},\ }\href
  {https://doi.org/10.1103/PhysRevLett.99.246403} {\bibfield  {journal}
  {\bibinfo  {journal} {Phys. Rev. Lett.}\ }\textbf {\bibinfo {volume} {99}},\
  \bibinfo {pages} {246403} (\bibinfo {year} {2007})}\BibitemShut {NoStop}%
\bibitem [{\citenamefont {Sakakibara}\ \emph {et~al.}(2020)\citenamefont
  {Sakakibara}, \citenamefont {Kotani}, \citenamefont {Obata},\ and\
  \citenamefont {Oda}}]{sakakibara2020prb}%
  \BibitemOpen
  \bibfield  {author} {\bibinfo {author} {\bibfnamefont {H.}~\bibnamefont
  {Sakakibara}}, \bibinfo {author} {\bibfnamefont {T.}~\bibnamefont {Kotani}},
  \bibinfo {author} {\bibfnamefont {M.}~\bibnamefont {Obata}},\ and\ \bibinfo
  {author} {\bibfnamefont {T.}~\bibnamefont {Oda}},\ }\bibfield  {title}
  {\bibinfo {title} {Finite electric-field approach to evaluate the vertex
  correction for the screened coulomb interaction in the quasiparticle
  self-consistent {$GW$} method},\ }\href
  {https://doi.org/10.1103/PhysRevB.101.205120} {\bibfield  {journal} {\bibinfo
   {journal} {Phys. Rev. B}\ }\textbf {\bibinfo {volume} {101}},\ \bibinfo
  {pages} {205120} (\bibinfo {year} {2020})}\BibitemShut {NoStop}%
\bibitem [{\citenamefont {{van Schilfgaarde}}\ \emph
  {et~al.}(2006)\citenamefont {{van Schilfgaarde}}, \citenamefont {Kotani},\
  and\ \citenamefont {Faleev}}]{van-schilfgaarde2006prl}%
  \BibitemOpen
  \bibfield  {author} {\bibinfo {author} {\bibfnamefont {M.}~\bibnamefont {{van
  Schilfgaarde}}}, \bibinfo {author} {\bibfnamefont {T.}~\bibnamefont
  {Kotani}},\ and\ \bibinfo {author} {\bibfnamefont {S.}~\bibnamefont
  {Faleev}},\ }\bibfield  {title} {\bibinfo {title} {{Quasiparticle
  Self-Consistent $GW$ Theory}},\ }\href@noop {} {\bibfield  {journal}
  {\bibinfo  {journal} {Phys. Rev. Lett.}\ }\textbf {\bibinfo {volume} {96}},\
  \bibinfo {pages} {226402} (\bibinfo {year} {2006})}\BibitemShut {NoStop}%
\bibitem [{\citenamefont {Botti}\ and\ \citenamefont
  {Marques}(2013)}]{botti2013prl}%
  \BibitemOpen
  \bibfield  {author} {\bibinfo {author} {\bibfnamefont {S.}~\bibnamefont
  {Botti}}\ and\ \bibinfo {author} {\bibfnamefont {M.~A.~L.}\ \bibnamefont
  {Marques}},\ }\bibfield  {title} {\bibinfo {title} {{Strong Renormalization
  of the Electronic Band Gap due to Lattice Polarization in the $GW$
  Formalism}},\ }\href@noop {} {\bibfield  {journal} {\bibinfo  {journal}
  {Phys. Rev. Lett.}\ }\textbf {\bibinfo {volume} {110}},\ \bibinfo {pages}
  {226404} (\bibinfo {year} {2013})}\BibitemShut {NoStop}%
\bibitem [{\citenamefont {Bhandari}\ \emph {et~al.}(2018)\citenamefont
  {Bhandari}, \citenamefont {van Schilgaarde}, \citenamefont {Kotani},\ and\
  \citenamefont {Lambrecht}}]{bhandari2018prm}%
  \BibitemOpen
  \bibfield  {author} {\bibinfo {author} {\bibfnamefont {C.}~\bibnamefont
  {Bhandari}}, \bibinfo {author} {\bibfnamefont {M.}~\bibnamefont {van
  Schilgaarde}}, \bibinfo {author} {\bibfnamefont {T.}~\bibnamefont {Kotani}},\
  and\ \bibinfo {author} {\bibfnamefont {W.~R.~L.}\ \bibnamefont {Lambrecht}},\
  }\bibfield  {title} {\bibinfo {title} {All-electron quasiparticle
  self-consistent {$GW$} band structures for {SrTiO$_{3}$} including lattice
  polarization corrections in different phases},\ }\href
  {https://doi.org/10.1103/PhysRevMaterials.2.013807} {\bibfield  {journal}
  {\bibinfo  {journal} {Phys. Rev. Materials}\ }\textbf {\bibinfo {volume}
  {2}},\ \bibinfo {pages} {013807} (\bibinfo {year} {2018})}\BibitemShut
  {NoStop}%
\bibitem [{\citenamefont {Chantis}\ \emph {et~al.}(2006)\citenamefont
  {Chantis}, \citenamefont {van Schilfgaarde},\ and\ \citenamefont
  {Kotani}}]{chantis2006prl}%
  \BibitemOpen
  \bibfield  {author} {\bibinfo {author} {\bibfnamefont {A.~N.}\ \bibnamefont
  {Chantis}}, \bibinfo {author} {\bibfnamefont {M.}~\bibnamefont {van
  Schilfgaarde}},\ and\ \bibinfo {author} {\bibfnamefont {T.}~\bibnamefont
  {Kotani}},\ }\bibfield  {title} {\bibinfo {title} {{Ab Initio Prediction of
  Conduction Band Spin Splitting in Zinc Blende Semiconductors}},\ }\href@noop
  {} {\bibfield  {journal} {\bibinfo  {journal} {Phys. Rev. Lett.}\ }\textbf
  {\bibinfo {volume} {96}},\ \bibinfo {pages} {086405} (\bibinfo {year}
  {2006})}\BibitemShut {NoStop}%
\bibitem [{\citenamefont {He}\ and\ \citenamefont
  {Franchini}(2012)}]{he2019prb}%
  \BibitemOpen
  \bibfield  {author} {\bibinfo {author} {\bibfnamefont {J.}~\bibnamefont
  {He}}\ and\ \bibinfo {author} {\bibfnamefont {C.}~\bibnamefont {Franchini}},\
  }\bibfield  {title} {\bibinfo {title} {Screened hybrid functional applied to
  $3d^0\rightarrow3d^8$ transition-metal perovskites {La$M$O$_3$}
  ({$M$}={Sc}--{Cu}): influence of the exchange mixing parameter on the
  structural, electronic and magnetic properties},\ }\href
  {https://doi.org/10.1103/PhysRevB.86.235117} {\bibfield  {journal} {\bibinfo
  {journal} {Physical Review B}\ }\textbf {\bibinfo {volume} {86}},\ \bibinfo
  {pages} {235117} (\bibinfo {year} {2012})},\ \bibinfo {note} {arXiv:
  1209.0486}\BibitemShut {NoStop}%
\bibitem [{\citenamefont {Mostofi}\ \emph {et~al.}(2008)\citenamefont
  {Mostofi}, \citenamefont {Yates}, \citenamefont {Lee}, \citenamefont {Souza},
  \citenamefont {Vanderbilt},\ and\ \citenamefont {Marzari}}]{mostofi2008cpc}%
  \BibitemOpen
  \bibfield  {author} {\bibinfo {author} {\bibfnamefont {A.~A.}\ \bibnamefont
  {Mostofi}}, \bibinfo {author} {\bibfnamefont {J.~R.}\ \bibnamefont {Yates}},
  \bibinfo {author} {\bibfnamefont {Y.-S.}\ \bibnamefont {Lee}}, \bibinfo
  {author} {\bibfnamefont {I.}~\bibnamefont {Souza}}, \bibinfo {author}
  {\bibfnamefont {D.}~\bibnamefont {Vanderbilt}},\ and\ \bibinfo {author}
  {\bibfnamefont {N.}~\bibnamefont {Marzari}},\ }\bibfield  {title} {\bibinfo
  {title} {{$\textsc{wannier90}$: A tool for obtaining maximally-localised
  Wannier functions}},\ }\href
  {https://doi.org/https://doi.org/10.1016/j.cpc.2007.11.016} {\bibfield
  {journal} {\bibinfo  {journal} {Computer Physics Communications}\ }\textbf
  {\bibinfo {volume} {178}},\ \bibinfo {pages} {685 } (\bibinfo {year}
  {2008})}\BibitemShut {NoStop}%
\bibitem [{\citenamefont {Mostofi}\ \emph {et~al.}(2014)\citenamefont
  {Mostofi}, \citenamefont {Yates}, \citenamefont {Pizzi}, \citenamefont {Lee},
  \citenamefont {Souza}, \citenamefont {Vanderbilt},\ and\ \citenamefont
  {Marzari}}]{mostofi2014cpc}%
  \BibitemOpen
  \bibfield  {author} {\bibinfo {author} {\bibfnamefont {A.~A.}\ \bibnamefont
  {Mostofi}}, \bibinfo {author} {\bibfnamefont {J.~R.}\ \bibnamefont {Yates}},
  \bibinfo {author} {\bibfnamefont {G.}~\bibnamefont {Pizzi}}, \bibinfo
  {author} {\bibfnamefont {Y.-S.}\ \bibnamefont {Lee}}, \bibinfo {author}
  {\bibfnamefont {I.}~\bibnamefont {Souza}}, \bibinfo {author} {\bibfnamefont
  {D.}~\bibnamefont {Vanderbilt}},\ and\ \bibinfo {author} {\bibfnamefont
  {N.}~\bibnamefont {Marzari}},\ }\bibfield  {title} {\bibinfo {title} {{An
  updated version of \textsc{wannier90}: A tool for obtaining
  maximally-localised Wannier functions}},\ }\href
  {https://doi.org/https://doi.org/10.1016/j.cpc.2014.05.003} {\bibfield
  {journal} {\bibinfo  {journal} {Computer Physics Communications}\ }\textbf
  {\bibinfo {volume} {185}},\ \bibinfo {pages} {2309 } (\bibinfo {year}
  {2014})}\BibitemShut {NoStop}%
\bibitem [{\citenamefont {McGuire}\ \emph {et~al.}(2015)\citenamefont
  {McGuire}, \citenamefont {Dixit}, \citenamefont {Cooper},\ and\ \citenamefont
  {Sales}}]{mcguire2015cm}%
  \BibitemOpen
  \bibfield  {author} {\bibinfo {author} {\bibfnamefont {M.~A.}\ \bibnamefont
  {McGuire}}, \bibinfo {author} {\bibfnamefont {H.}~\bibnamefont {Dixit}},
  \bibinfo {author} {\bibfnamefont {V.~R.}\ \bibnamefont {Cooper}},\ and\
  \bibinfo {author} {\bibfnamefont {B.~C.}\ \bibnamefont {Sales}},\ }\bibfield
  {title} {\bibinfo {title} {{Coupling of Crystal Structure and Magnetism in
  the Layered, Ferromagnetic Insulator CrI$_3$}},\ }\href@noop {} {\bibfield
  {journal} {\bibinfo  {journal} {Chemistry of Materials}\ }\textbf {\bibinfo
  {volume} {27}},\ \bibinfo {pages} {612} (\bibinfo {year} {2015})}\BibitemShut
  {NoStop}%
\bibitem [{\citenamefont {Carteaux}\ \emph {et~al.}(1995)\citenamefont
  {Carteaux}, \citenamefont {Brunet}, \citenamefont {Ouvrard},\ and\
  \citenamefont {Andre}}]{carteaux1995jpcm}%
  \BibitemOpen
  \bibfield  {author} {\bibinfo {author} {\bibfnamefont {V.}~\bibnamefont
  {Carteaux}}, \bibinfo {author} {\bibfnamefont {D.}~\bibnamefont {Brunet}},
  \bibinfo {author} {\bibfnamefont {G.}~\bibnamefont {Ouvrard}},\ and\ \bibinfo
  {author} {\bibfnamefont {G.}~\bibnamefont {Andre}},\ }\bibfield  {title}
  {\bibinfo {title} {{Crystallographic, magnetic and electronic structures of a
  new layered ferromagnetic compound Cr$_2$Ge$_2$Te$_6$}},\ }\href
  {https://doi.org/10.1088/0953-8984/7/1/008} {\bibfield  {journal} {\bibinfo
  {journal} {Journal of Physics: Condensed Matter}\ }\textbf {\bibinfo {volume}
  {7}},\ \bibinfo {pages} {69} (\bibinfo {year} {1995})}\BibitemShut {NoStop}%
\bibitem [{\citenamefont {Deiseroth}\ \emph {et~al.}(2006)\citenamefont
  {Deiseroth}, \citenamefont {Aleksandrov}, \citenamefont {Reiner},
  \citenamefont {Kienle},\ and\ \citenamefont {Kremer}}]{deiseroth2006ejic}%
  \BibitemOpen
  \bibfield  {author} {\bibinfo {author} {\bibfnamefont {H.-J.}\ \bibnamefont
  {Deiseroth}}, \bibinfo {author} {\bibfnamefont {K.}~\bibnamefont
  {Aleksandrov}}, \bibinfo {author} {\bibfnamefont {C.}~\bibnamefont {Reiner}},
  \bibinfo {author} {\bibfnamefont {L.}~\bibnamefont {Kienle}},\ and\ \bibinfo
  {author} {\bibfnamefont {R.~K.}\ \bibnamefont {Kremer}},\ }\bibfield  {title}
  {\bibinfo {title} {{Fe$_3$GeTe$_2$ and Ni$_3$GeTe$_2$ – Two New Layered
  Transition-Metal Compounds: Crystal Structures, HRTEM Investigations, and
  Magnetic and Electrical Properties}},\ }\href
  {https://doi.org/10.1002/ejic.200501020} {\bibfield  {journal} {\bibinfo
  {journal} {European Journal of Inorganic Chemistry}\ }\textbf {\bibinfo
  {volume} {2006}},\ \bibinfo {pages} {1561} (\bibinfo {year}
  {2006})}\BibitemShut {NoStop}%
\bibitem [{\citenamefont {Liechtenstein}\ \emph {et~al.}(1995)\citenamefont
  {Liechtenstein}, \citenamefont {Anisimov},\ and\ \citenamefont
  {Zaanen}}]{liechtenstein1995prb}%
  \BibitemOpen
  \bibfield  {author} {\bibinfo {author} {\bibfnamefont {A.~I.}\ \bibnamefont
  {Liechtenstein}}, \bibinfo {author} {\bibfnamefont {V.~I.}\ \bibnamefont
  {Anisimov}},\ and\ \bibinfo {author} {\bibfnamefont {J.}~\bibnamefont
  {Zaanen}},\ }\bibfield  {title} {\bibinfo {title} {Density-functional theory
  and strong interactions: {Orbital} ordering in {Mott-Hubbard} insulators},\
  }\href@noop {} {\bibfield  {journal} {\bibinfo  {journal} {Phys. Rev. B}\
  }\textbf {\bibinfo {volume} {52}},\ \bibinfo {pages} {R5467} (\bibinfo {year}
  {1995})}\BibitemShut {NoStop}%
\bibitem [{\citenamefont {Petukhov}\ \emph {et~al.}(2003)\citenamefont
  {Petukhov}, \citenamefont {Mazin}, \citenamefont {Chioncel},\ and\
  \citenamefont {Lichtenstein}}]{petukhov2003prb}%
  \BibitemOpen
  \bibfield  {author} {\bibinfo {author} {\bibfnamefont {A.~G.}\ \bibnamefont
  {Petukhov}}, \bibinfo {author} {\bibfnamefont {I.~I.}\ \bibnamefont {Mazin}},
  \bibinfo {author} {\bibfnamefont {L.}~\bibnamefont {Chioncel}},\ and\
  \bibinfo {author} {\bibfnamefont {A.~I.}\ \bibnamefont {Lichtenstein}},\
  }\bibfield  {title} {\bibinfo {title} {{Correlated metals and the LDA$+U$
  method}},\ }\href {https://doi.org/10.1103/PhysRevB.67.153106} {\bibfield
  {journal} {\bibinfo  {journal} {Phys. Rev. B}\ }\textbf {\bibinfo {volume}
  {67}},\ \bibinfo {pages} {153106} (\bibinfo {year} {2003})}\BibitemShut
  {NoStop}%
\bibitem [{\citenamefont {Perdew}\ \emph {et~al.}(1996)\citenamefont {Perdew},
  \citenamefont {Burke},\ and\ \citenamefont {Ernzerhof}}]{perdew1996prl}%
  \BibitemOpen
  \bibfield  {author} {\bibinfo {author} {\bibfnamefont {J.~P.}\ \bibnamefont
  {Perdew}}, \bibinfo {author} {\bibfnamefont {K.}~\bibnamefont {Burke}},\ and\
  \bibinfo {author} {\bibfnamefont {M.}~\bibnamefont {Ernzerhof}},\ }\bibfield
  {title} {\bibinfo {title} {{Generalized Gradient Approximation Made
  Simple}},\ }\href {https://doi.org/10.1103/PhysRevLett.77.3865} {\bibfield
  {journal} {\bibinfo  {journal} {Phys. Rev. Lett.}\ }\textbf {\bibinfo
  {volume} {77}},\ \bibinfo {pages} {3865} (\bibinfo {year}
  {1996})}\BibitemShut {NoStop}%
\bibitem [{\citenamefont {Kotani}\ and\ \citenamefont
  {Kino}(2009)}]{kotani2009jpcm}%
  \BibitemOpen
  \bibfield  {author} {\bibinfo {author} {\bibfnamefont {T.}~\bibnamefont
  {Kotani}}\ and\ \bibinfo {author} {\bibfnamefont {H.}~\bibnamefont {Kino}},\
  }\bibfield  {title} {\bibinfo {title} {Re-examination of half-metallic
  ferromagnetism for doped {LaMnO$_3$} in a quasiparticle self-consistent
  {$GW$} method},\ }\href@noop {} {\bibfield  {journal} {\bibinfo  {journal}
  {Journal of Physics: Condensed Matter}\ }\textbf {\bibinfo {volume} {21}},\
  \bibinfo {pages} {266002} (\bibinfo {year} {2009})}\BibitemShut {NoStop}%
\bibitem [{\citenamefont {Jang}\ \emph {et~al.}(2015)\citenamefont {Jang},
  \citenamefont {Kotani}, \citenamefont {Kino}, \citenamefont {Kuroki},\ and\
  \citenamefont {Han}}]{jang2015sr}%
  \BibitemOpen
  \bibfield  {author} {\bibinfo {author} {\bibfnamefont {S.~W.}\ \bibnamefont
  {Jang}}, \bibinfo {author} {\bibfnamefont {T.}~\bibnamefont {Kotani}},
  \bibinfo {author} {\bibfnamefont {H.}~\bibnamefont {Kino}}, \bibinfo {author}
  {\bibfnamefont {K.}~\bibnamefont {Kuroki}},\ and\ \bibinfo {author}
  {\bibfnamefont {M.~J.}\ \bibnamefont {Han}},\ }\bibfield  {title} {\bibinfo
  {title} {{Quasiparticle self-consistent $GW$ study of cuprates: electronic
  structure, model parameters, and the two-band theory for $T_\text{C}$}},\
  }\href {https://doi.org/10.1038/srep12050} {\bibfield  {journal} {\bibinfo
  {journal} {Scientific Reports}\ }\textbf {\bibinfo {volume} {5}},\ \bibinfo
  {pages} {12050} (\bibinfo {year} {2015})}\BibitemShut {NoStop}%
\bibitem [{sm()}]{sm}%
  \BibitemOpen
  \href@noop {} {}\bibinfo {note} {See Supplemental Material for more detailed
  band structures and density of states calculated using various
  methods.}\BibitemShut {Stop}%
\bibitem [{\citenamefont {Janthon}\ \emph {et~al.}(2014)\citenamefont
  {Janthon}, \citenamefont {Luo}, \citenamefont {Kozlov}, \citenamefont
  {Viñes}, \citenamefont {Limtrakul}, \citenamefont {Truhlar},\ and\
  \citenamefont {Illas}}]{patanachai2014jctc}%
  \BibitemOpen
  \bibfield  {author} {\bibinfo {author} {\bibfnamefont {P.}~\bibnamefont
  {Janthon}}, \bibinfo {author} {\bibfnamefont {S.~A.}\ \bibnamefont {Luo}},
  \bibinfo {author} {\bibfnamefont {S.~M.}\ \bibnamefont {Kozlov}}, \bibinfo
  {author} {\bibfnamefont {F.}~\bibnamefont {Viñes}}, \bibinfo {author}
  {\bibfnamefont {J.}~\bibnamefont {Limtrakul}}, \bibinfo {author}
  {\bibfnamefont {D.~G.}\ \bibnamefont {Truhlar}},\ and\ \bibinfo {author}
  {\bibfnamefont {F.}~\bibnamefont {Illas}},\ }\bibfield  {title} {\bibinfo
  {title} {{Bulk Properties of Transition Metals: A Challenge for the Design of
  Universal Density Functionals}},\ }\href@noop {} {\bibfield  {journal}
  {\bibinfo  {journal} {Journal of Chemical Theory and Computation}\ }\textbf
  {\bibinfo {volume} {10}},\ \bibinfo {pages} {3832} (\bibinfo {year}
  {2014})}\BibitemShut {NoStop}%
\bibitem [{\citenamefont {Meng}\ \emph {et~al.}(2016)\citenamefont {Meng},
  \citenamefont {Liu}, \citenamefont {Huo}, \citenamefont {Guo}, \citenamefont
  {Cao}, \citenamefont {Peng}, \citenamefont {Dearden}, \citenamefont {Gonze},
  \citenamefont {Yang}, \citenamefont {Wang}, \citenamefont {Jiao},
  \citenamefont {Li},\ and\ \citenamefont {Wen}}]{yu2016jctc}%
  \BibitemOpen
  \bibfield  {author} {\bibinfo {author} {\bibfnamefont {Y.}~\bibnamefont
  {Meng}}, \bibinfo {author} {\bibfnamefont {X.-W.}\ \bibnamefont {Liu}},
  \bibinfo {author} {\bibfnamefont {C.-F.}\ \bibnamefont {Huo}}, \bibinfo
  {author} {\bibfnamefont {W.-P.}\ \bibnamefont {Guo}}, \bibinfo {author}
  {\bibfnamefont {D.-B.}\ \bibnamefont {Cao}}, \bibinfo {author} {\bibfnamefont
  {Q.}~\bibnamefont {Peng}}, \bibinfo {author} {\bibfnamefont {A.}~\bibnamefont
  {Dearden}}, \bibinfo {author} {\bibfnamefont {X.}~\bibnamefont {Gonze}},
  \bibinfo {author} {\bibfnamefont {Y.}~\bibnamefont {Yang}}, \bibinfo {author}
  {\bibfnamefont {J.}~\bibnamefont {Wang}}, \bibinfo {author} {\bibfnamefont
  {H.}~\bibnamefont {Jiao}}, \bibinfo {author} {\bibfnamefont {Y.}~\bibnamefont
  {Li}},\ and\ \bibinfo {author} {\bibfnamefont {X.-D.}\ \bibnamefont {Wen}},\
  }\bibfield  {title} {\bibinfo {title} {{When Density Functional
  Approximations Meet Iron Oxides}},\ }\href@noop {} {\bibfield  {journal}
  {\bibinfo  {journal} {Journal of Chemical Theory and Computation}\ }\textbf
  {\bibinfo {volume} {12}},\ \bibinfo {pages} {5132} (\bibinfo {year}
  {2016})}\BibitemShut {NoStop}%
\bibitem [{\citenamefont {Wang}\ \emph {et~al.}(2011)\citenamefont {Wang},
  \citenamefont {Eyert},\ and\ \citenamefont
  {Schwingenschlögl}}]{wang2011jpcm}%
  \BibitemOpen
  \bibfield  {author} {\bibinfo {author} {\bibfnamefont {H.}~\bibnamefont
  {Wang}}, \bibinfo {author} {\bibfnamefont {V.}~\bibnamefont {Eyert}},\ and\
  \bibinfo {author} {\bibfnamefont {U.}~\bibnamefont {Schwingenschlögl}},\
  }\bibfield  {title} {\bibinfo {title} {{Electronic structure and magnetic
  ordering of the semiconducting chromium trihalides CrCl$_3$, CrBr$_3$, and
  CrI$_3$}},\ }\href {https://doi.org/10.1088/0953-8984/23/11/116003}
  {\bibfield  {journal} {\bibinfo  {journal} {Journal of Physics: Condensed
  Matter}\ }\textbf {\bibinfo {volume} {23}},\ \bibinfo {pages} {116003}
  (\bibinfo {year} {2011})}\BibitemShut {NoStop}%
\bibitem [{\citenamefont {Dillon}\ and\ \citenamefont
  {Olson}(1965)}]{dillon1965jap}%
  \BibitemOpen
  \bibfield  {author} {\bibinfo {author} {\bibfnamefont {J.~F.}\ \bibnamefont
  {Dillon}}\ and\ \bibinfo {author} {\bibfnamefont {C.~E.}\ \bibnamefont
  {Olson}},\ }\bibfield  {title} {\bibinfo {title} {{Magnetization, Resonance,
  and Optical Properties of the Ferromagnet CrI$_3$}},\ }\href
  {https://doi.org/10.1063/1.1714194} {\bibfield  {journal} {\bibinfo
  {journal} {Journal of Applied Physics}\ }\textbf {\bibinfo {volume} {36}},\
  \bibinfo {pages} {1259} (\bibinfo {year} {1965})}\BibitemShut {NoStop}%
\bibitem [{\citenamefont {Li}\ \emph {et~al.}(2018)\citenamefont {Li},
  \citenamefont {Wang}, \citenamefont {Guo}, \citenamefont {Gu}, \citenamefont
  {Sun}, \citenamefont {He}, \citenamefont {Zhou}, \citenamefont {Gu},
  \citenamefont {Nie},\ and\ \citenamefont {Pan}}]{li2018prb}%
  \BibitemOpen
  \bibfield  {author} {\bibinfo {author} {\bibfnamefont {Y.~F.}\ \bibnamefont
  {Li}}, \bibinfo {author} {\bibfnamefont {W.}~\bibnamefont {Wang}}, \bibinfo
  {author} {\bibfnamefont {W.}~\bibnamefont {Guo}}, \bibinfo {author}
  {\bibfnamefont {C.~Y.}\ \bibnamefont {Gu}}, \bibinfo {author} {\bibfnamefont
  {H.~Y.}\ \bibnamefont {Sun}}, \bibinfo {author} {\bibfnamefont
  {L.}~\bibnamefont {He}}, \bibinfo {author} {\bibfnamefont {J.}~\bibnamefont
  {Zhou}}, \bibinfo {author} {\bibfnamefont {Z.~B.}\ \bibnamefont {Gu}},
  \bibinfo {author} {\bibfnamefont {Y.~F.}\ \bibnamefont {Nie}},\ and\ \bibinfo
  {author} {\bibfnamefont {X.~Q.}\ \bibnamefont {Pan}},\ }\bibfield  {title}
  {\bibinfo {title} {Electronic structure of ferromagnetic semiconductor
  {CrGeTe$_3$} by angle-resolved photoemission spectroscopy},\ }\href
  {https://doi.org/10.1103/PhysRevB.98.125127} {\bibfield  {journal} {\bibinfo
  {journal} {Phys. Rev. B}\ }\textbf {\bibinfo {volume} {98}},\ \bibinfo
  {pages} {125127} (\bibinfo {year} {2018})}\BibitemShut {NoStop}%
\bibitem [{\citenamefont {Suzuki}\ \emph {et~al.}(2019)\citenamefont {Suzuki},
  \citenamefont {Gao}, \citenamefont {Koshiishi}, \citenamefont {Nakata},
  \citenamefont {Hagiwara}, \citenamefont {Lin}, \citenamefont {Wan},
  \citenamefont {Kumigashira}, \citenamefont {Ono}, \citenamefont {Kang},
  \citenamefont {Kang}, \citenamefont {Yu}, \citenamefont {Kobayashi},
  \citenamefont {Cheong},\ and\ \citenamefont {Fujimori}}]{suzuki2019prb}%
  \BibitemOpen
  \bibfield  {author} {\bibinfo {author} {\bibfnamefont {M.}~\bibnamefont
  {Suzuki}}, \bibinfo {author} {\bibfnamefont {B.}~\bibnamefont {Gao}},
  \bibinfo {author} {\bibfnamefont {K.}~\bibnamefont {Koshiishi}}, \bibinfo
  {author} {\bibfnamefont {S.}~\bibnamefont {Nakata}}, \bibinfo {author}
  {\bibfnamefont {K.}~\bibnamefont {Hagiwara}}, \bibinfo {author}
  {\bibfnamefont {C.}~\bibnamefont {Lin}}, \bibinfo {author} {\bibfnamefont
  {Y.~X.}\ \bibnamefont {Wan}}, \bibinfo {author} {\bibfnamefont
  {H.}~\bibnamefont {Kumigashira}}, \bibinfo {author} {\bibfnamefont
  {K.}~\bibnamefont {Ono}}, \bibinfo {author} {\bibfnamefont {S.}~\bibnamefont
  {Kang}}, \bibinfo {author} {\bibfnamefont {S.}~\bibnamefont {Kang}}, \bibinfo
  {author} {\bibfnamefont {J.}~\bibnamefont {Yu}}, \bibinfo {author}
  {\bibfnamefont {M.}~\bibnamefont {Kobayashi}}, \bibinfo {author}
  {\bibfnamefont {S.-W.}\ \bibnamefont {Cheong}},\ and\ \bibinfo {author}
  {\bibfnamefont {A.}~\bibnamefont {Fujimori}},\ }\bibfield  {title} {\bibinfo
  {title} {{Coulomb-interaction effect on the two-dimensional electronic
  structure of the van der {Waals} ferromagnet Cr$_{2}$Ge$_{2}$Te$_{6}$}},\
  }\href {https://doi.org/10.1103/PhysRevB.99.161401} {\bibfield  {journal}
  {\bibinfo  {journal} {Phys. Rev. B}\ }\textbf {\bibinfo {volume} {99}},\
  \bibinfo {pages} {161401} (\bibinfo {year} {2019})}\BibitemShut {NoStop}%
\bibitem [{\citenamefont {Ji}\ \emph {et~al.}(2013)\citenamefont {Ji},
  \citenamefont {Stokes}, \citenamefont {Alegria}, \citenamefont {Blomberg},
  \citenamefont {Tanatar}, \citenamefont {Reijnders}, \citenamefont {Schoop},
  \citenamefont {Liang}, \citenamefont {Prozorov}, \citenamefont {Burch},
  \citenamefont {Ong}, \citenamefont {Petta},\ and\ \citenamefont
  {Cava}}]{ji2013jap}%
  \BibitemOpen
  \bibfield  {author} {\bibinfo {author} {\bibfnamefont {H.}~\bibnamefont
  {Ji}}, \bibinfo {author} {\bibfnamefont {R.~A.}\ \bibnamefont {Stokes}},
  \bibinfo {author} {\bibfnamefont {L.~D.}\ \bibnamefont {Alegria}}, \bibinfo
  {author} {\bibfnamefont {E.~C.}\ \bibnamefont {Blomberg}}, \bibinfo {author}
  {\bibfnamefont {M.~A.}\ \bibnamefont {Tanatar}}, \bibinfo {author}
  {\bibfnamefont {A.}~\bibnamefont {Reijnders}}, \bibinfo {author}
  {\bibfnamefont {L.~M.}\ \bibnamefont {Schoop}}, \bibinfo {author}
  {\bibfnamefont {T.}~\bibnamefont {Liang}}, \bibinfo {author} {\bibfnamefont
  {R.}~\bibnamefont {Prozorov}}, \bibinfo {author} {\bibfnamefont {K.~S.}\
  \bibnamefont {Burch}}, \bibinfo {author} {\bibfnamefont {N.~P.}\ \bibnamefont
  {Ong}}, \bibinfo {author} {\bibfnamefont {J.~R.}\ \bibnamefont {Petta}},\
  and\ \bibinfo {author} {\bibfnamefont {R.~J.}\ \bibnamefont {Cava}},\
  }\bibfield  {title} {\bibinfo {title} {{A ferromagnetic insulating substrate
  for the epitaxial growth of topological insulators}},\ }\href
  {https://doi.org/10.1063/1.4822092} {\bibfield  {journal} {\bibinfo
  {journal} {Journal of Applied Physics}\ }\textbf {\bibinfo {volume} {114}},\
  \bibinfo {pages} {114907} (\bibinfo {year} {2013})}\BibitemShut {NoStop}%
\bibitem [{\citenamefont {Ricca}\ \emph {et~al.}(2020)\citenamefont {Ricca},
  \citenamefont {Timrov}, \citenamefont {Cococcioni}, \citenamefont {Marzari},\
  and\ \citenamefont {Aschauer}}]{ricca2020arxiv}%
  \BibitemOpen
  \bibfield  {author} {\bibinfo {author} {\bibfnamefont {C.}~\bibnamefont
  {Ricca}}, \bibinfo {author} {\bibfnamefont {I.}~\bibnamefont {Timrov}},
  \bibinfo {author} {\bibfnamefont {M.}~\bibnamefont {Cococcioni}}, \bibinfo
  {author} {\bibfnamefont {N.}~\bibnamefont {Marzari}},\ and\ \bibinfo {author}
  {\bibfnamefont {U.}~\bibnamefont {Aschauer}},\ }\href@noop {} {\bibinfo
  {title} {Self-consistent {DFT+$U$+$V$} study of oxygen vacancies in
  {SrTiO$_3$}}} (\bibinfo {year} {2020}),\ \Eprint
  {https://arxiv.org/abs/2001.06540} {arXiv:2001.06540 [cond-mat.mtrl-sci]}
  \BibitemShut {NoStop}%
\bibitem [{\citenamefont {Kutepov}(2019)}]{kutepov2019arxiv}%
  \BibitemOpen
  \bibfield  {author} {\bibinfo {author} {\bibfnamefont {A.~L.}\ \bibnamefont
  {Kutepov}},\ }\href@noop {} {\bibinfo {title} {Self-consistent {$GW$} method:
  {$O(N)$} algorithm for the polarizability and the self energy}} (\bibinfo
  {year} {2019}),\ \Eprint {https://arxiv.org/abs/1911.05633} {arXiv:1911.05633
  [cond-mat.mtrl-sci]} \BibitemShut {NoStop}%
\end{thebibliography}%
\bigskip

\newpage

\makeatletter
\renewcommand{\fnum@figure}{FIG. S\thefigure}
\makeatother
\setcounter{figure}{0} 

\section*{Supplemental}

\subsection{Band structures and Density of States calculated in DFT and QSGW}

\subsubsection{VI$_{3}$}

\begin{figure}[hbt]
	\centering
	\includegraphics[width=1.0\linewidth]{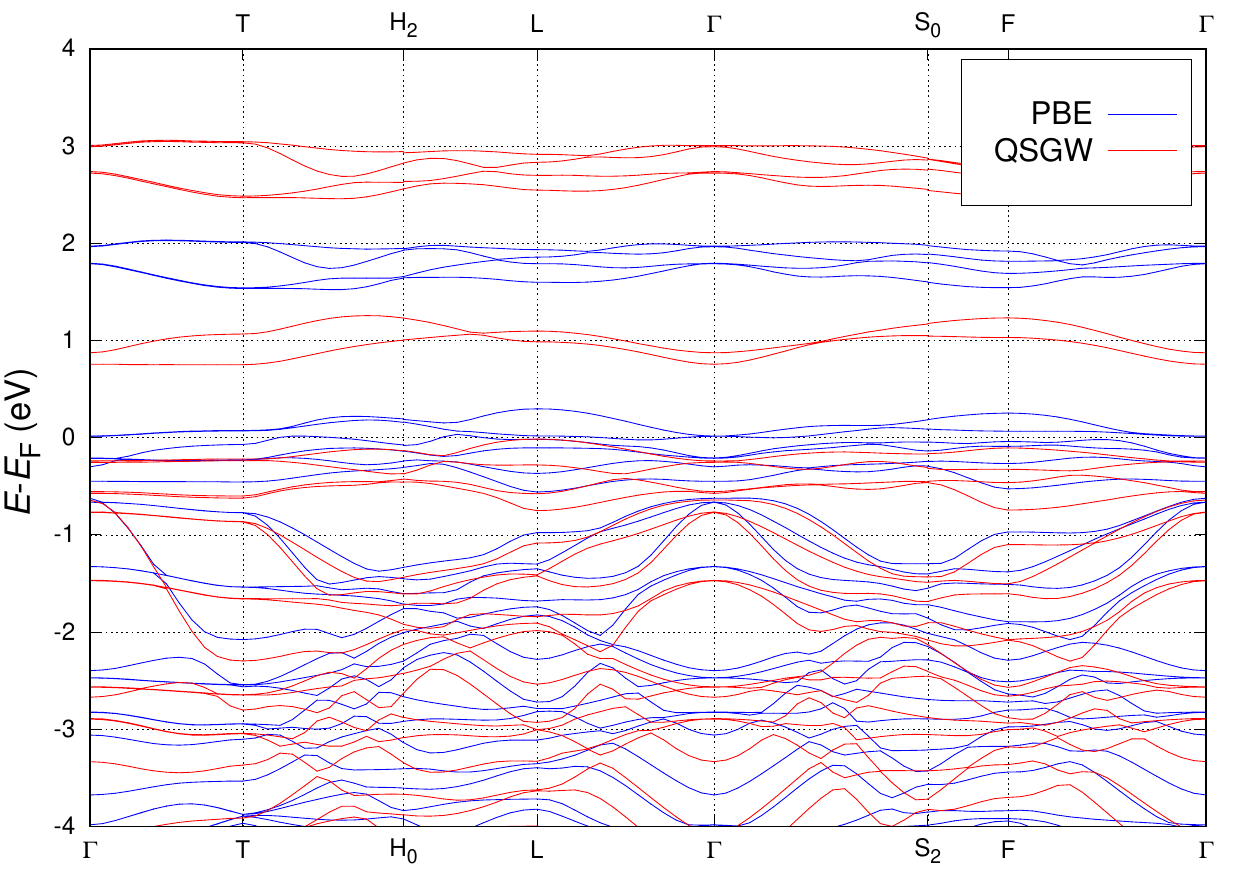} \\
	\includegraphics[width=1.0\linewidth]{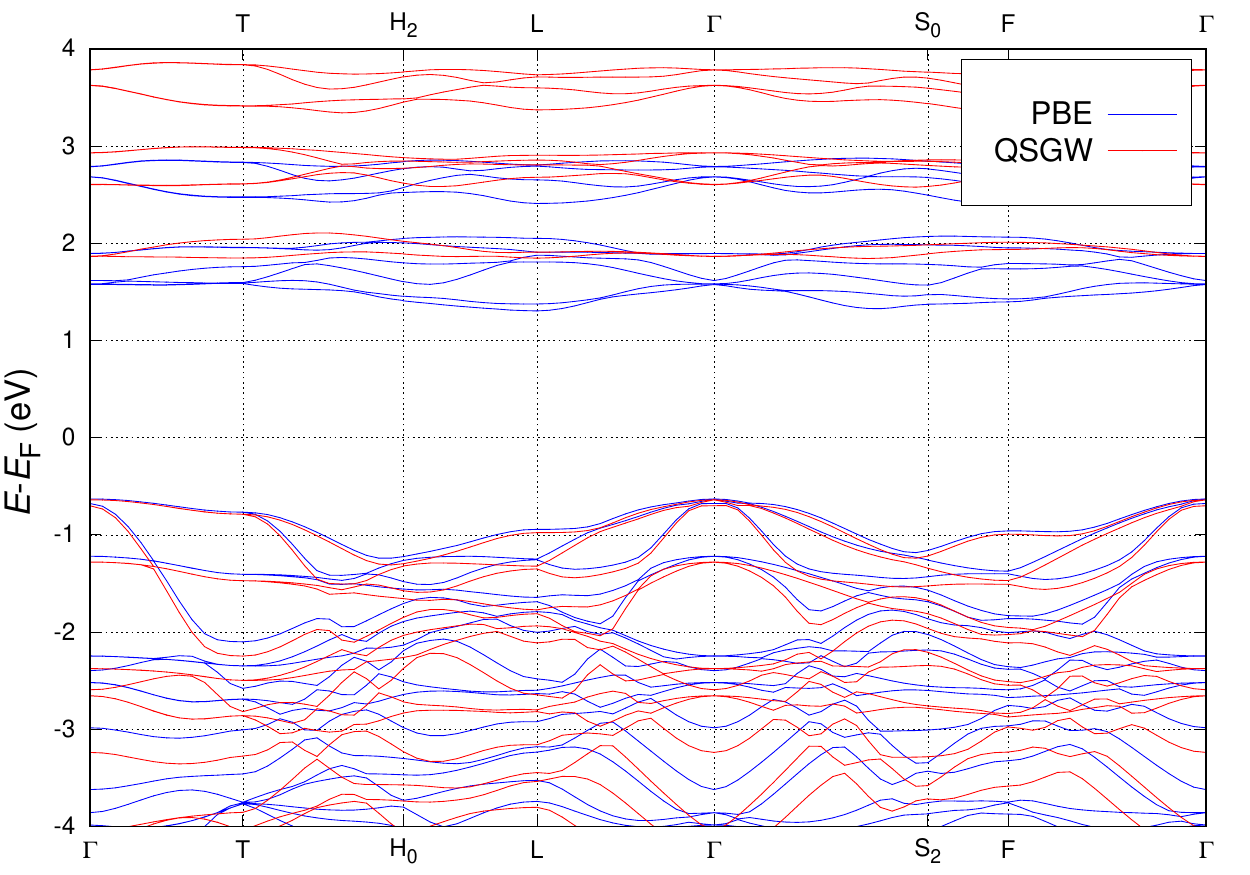} 
	\caption{Spin-polarized band structures of VI$_{3}$. 
          The top (bottom) panel shows the majority (minority) spin state.
          For comparison, QSGW (red) and PBE (blue) results are shown on the same panel.}
	\label{fig:vi3band}	
\end{figure}

\begin{figure}[hbt]
  \centering
  \vspace*{62pt}
	\includegraphics[width=1.0\linewidth]{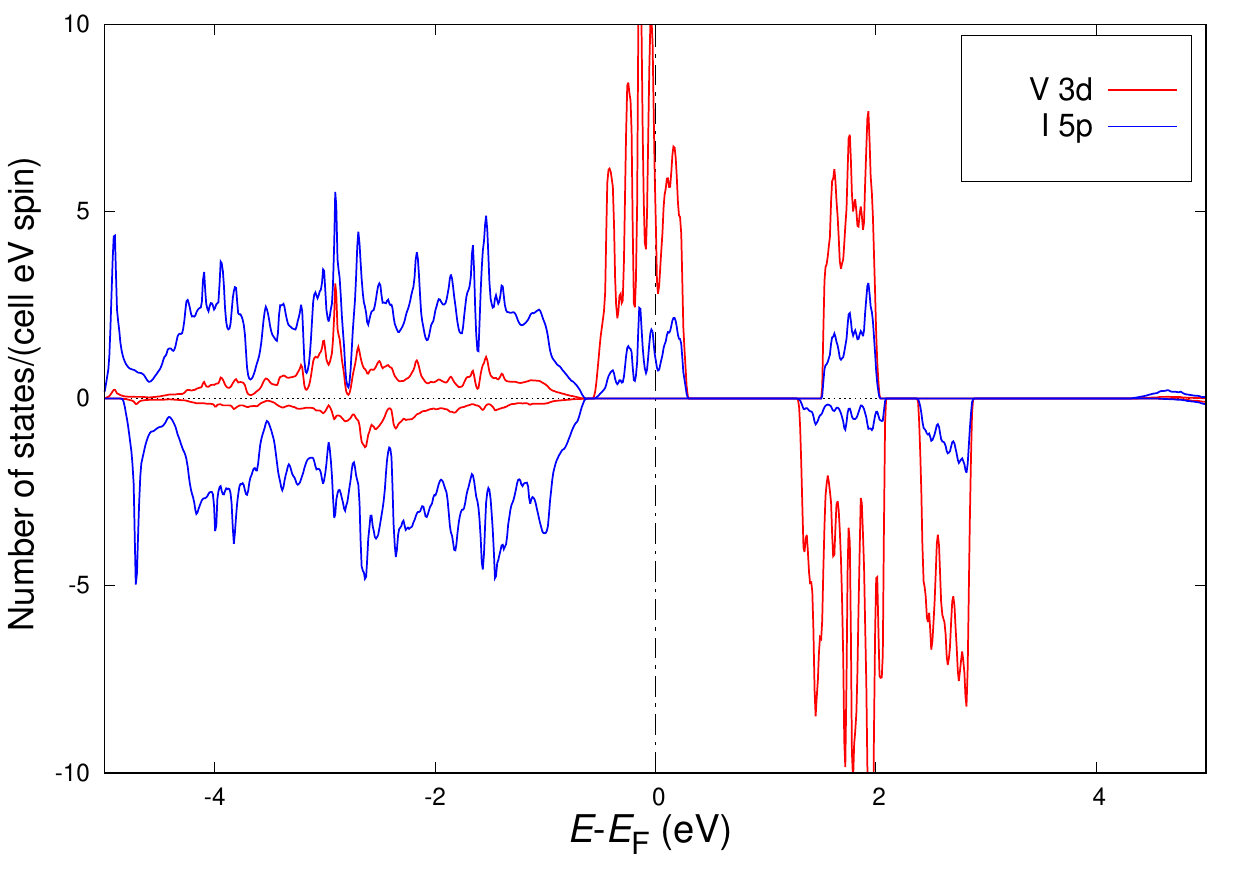} \\
	\includegraphics[width=1.0\linewidth]{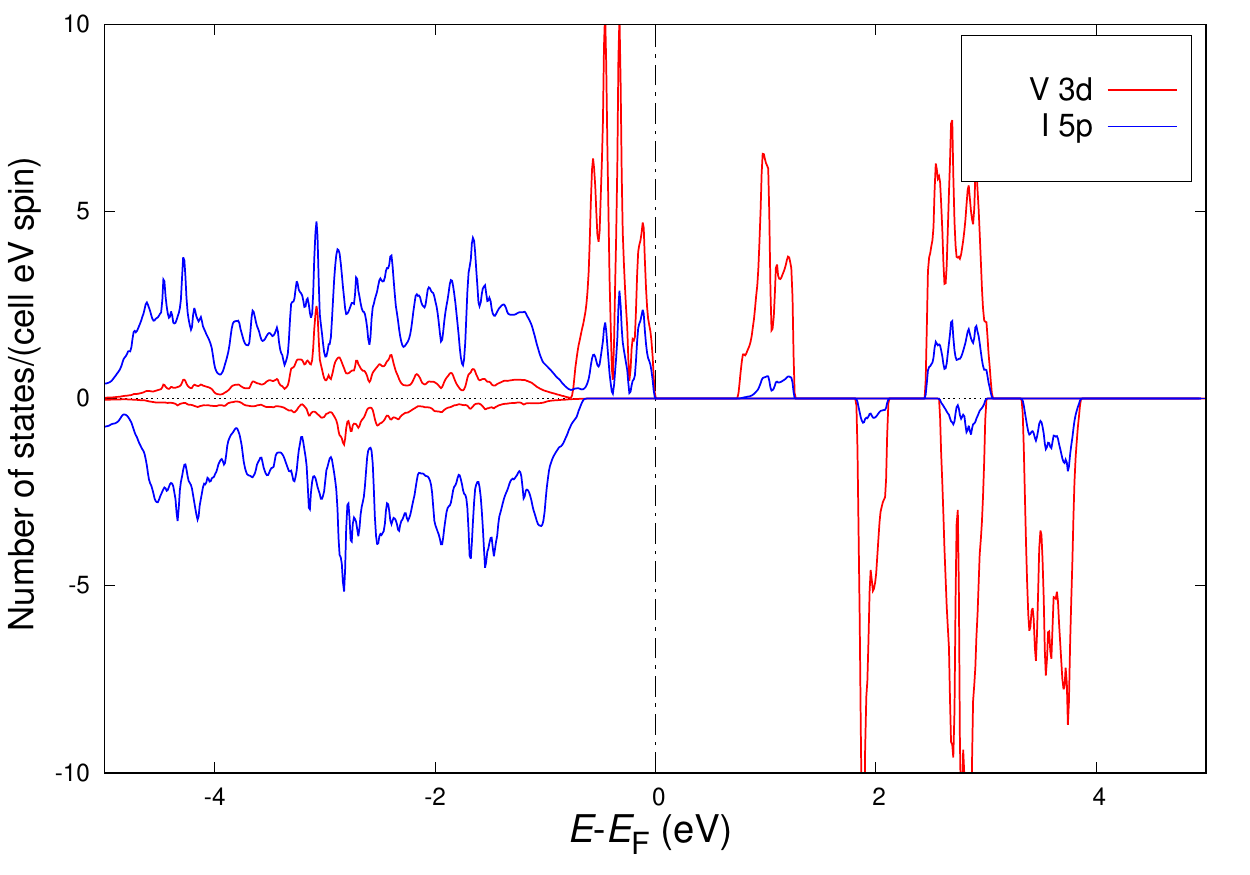}
	\caption{The partial density of states projected on V-$3d$ and I-$5p$ states in VI$_{3}$.
          The top (bottom) panel shows DFT-PBE (QSGW) results.
          DFT can not open a gap and result in a metallic state.
          Within QSGW, the minority spin channel has a much larger bandgap than the majority spin channel.}
	\label{fig:vi3pdos}	
\end{figure}

\clearpage

\subsubsection{CrGeTe$_{3}$}

\begin{figure}[hbt]
	\centering
	\includegraphics[width=1.0\linewidth]{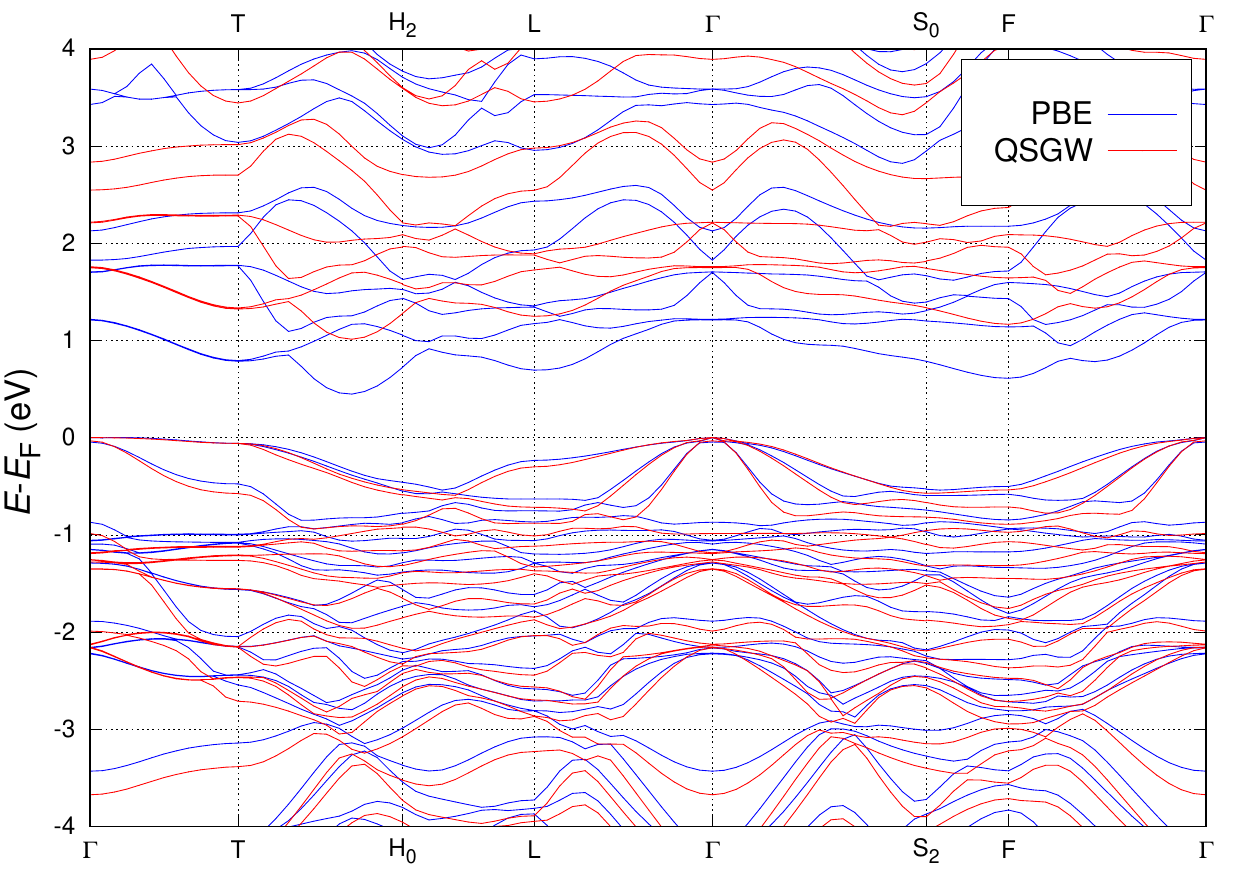} \\
	\includegraphics[width=1.0\linewidth]{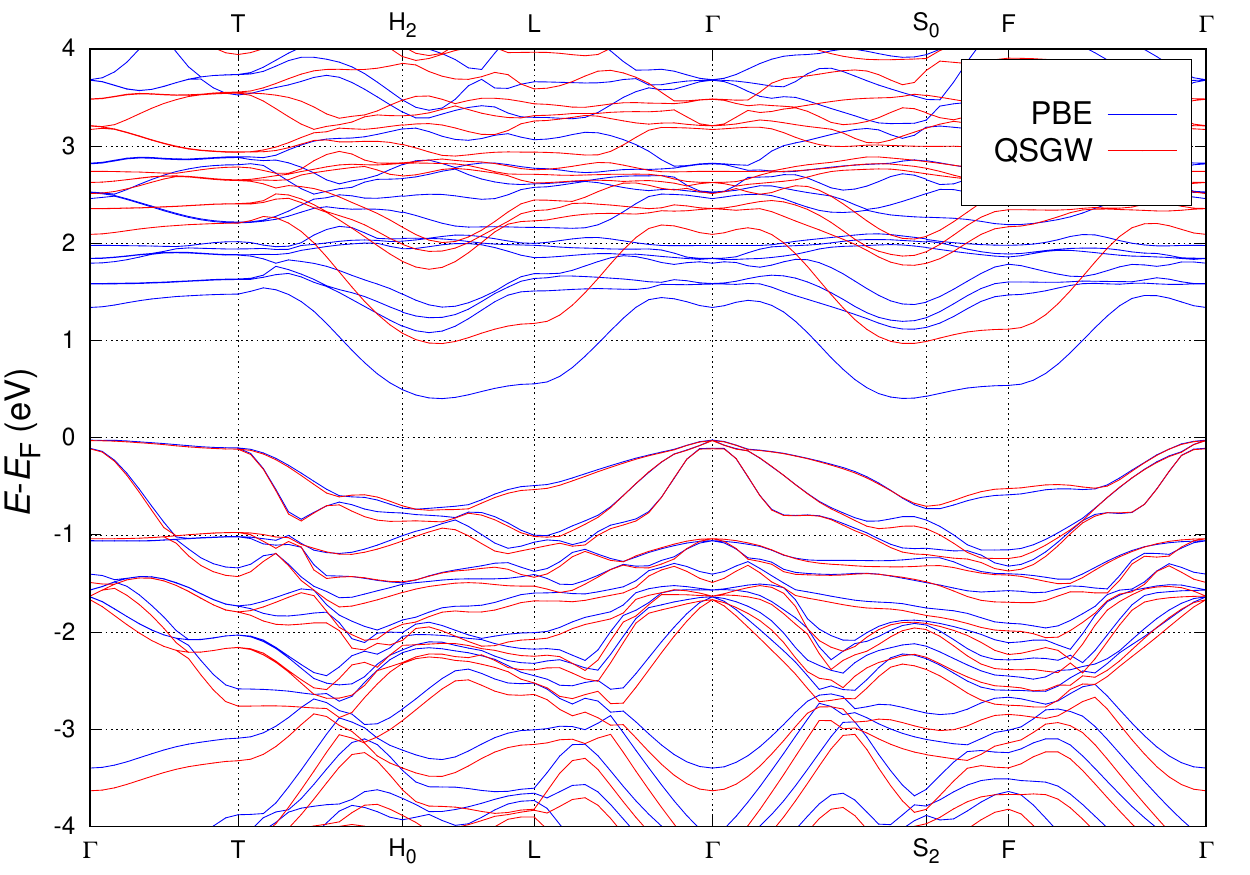}
	\caption{Spin-polarized band structures of CrGeTe$_{3}$ calculated in DFT and QSGW.
          The top (bottom) panel shows the majority (minority) spin state.}
	\label{fig:Cgt3band}	
\end{figure}

\begin{figure}[hbt]
  \centering
  \vspace*{40pt}
	\includegraphics[width=1.0\linewidth]{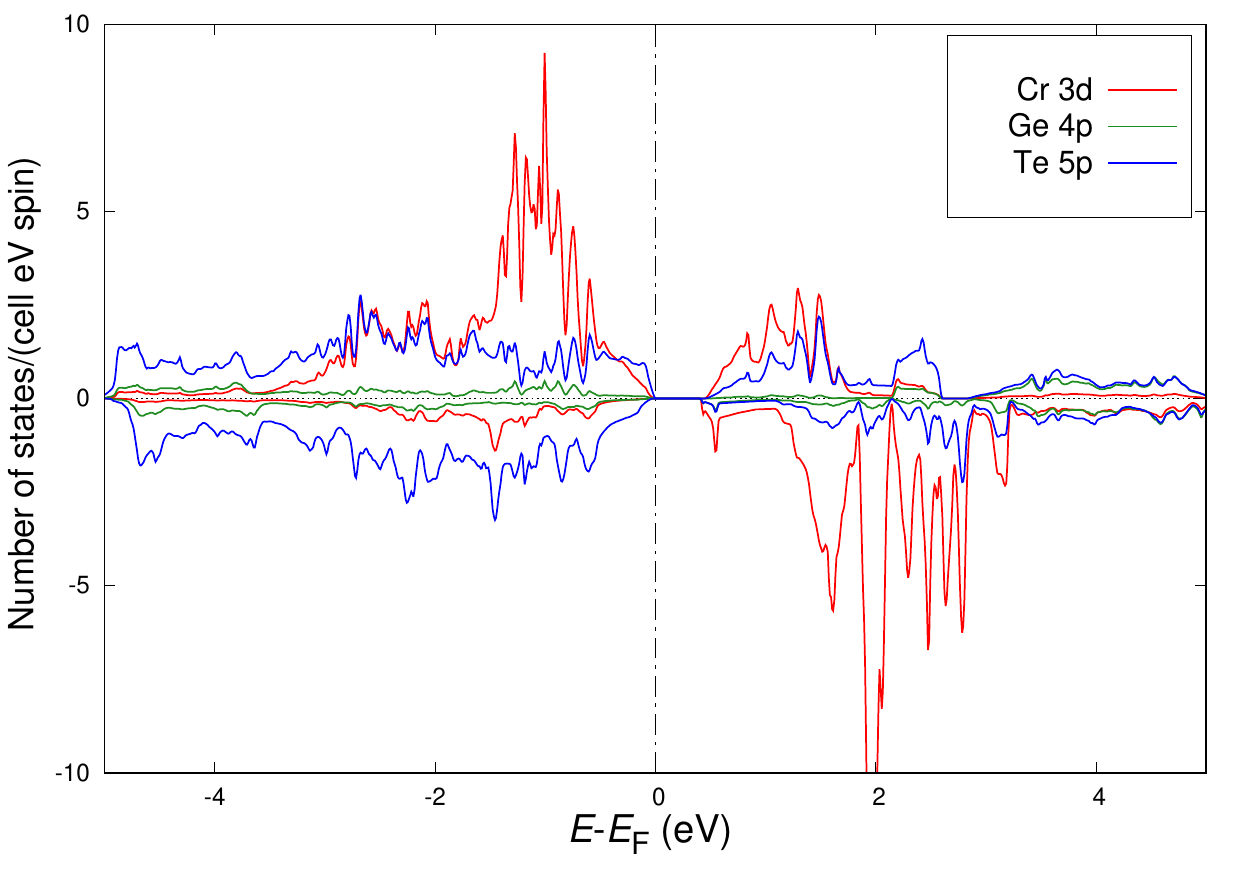} \\
	\includegraphics[width=1.0\linewidth]{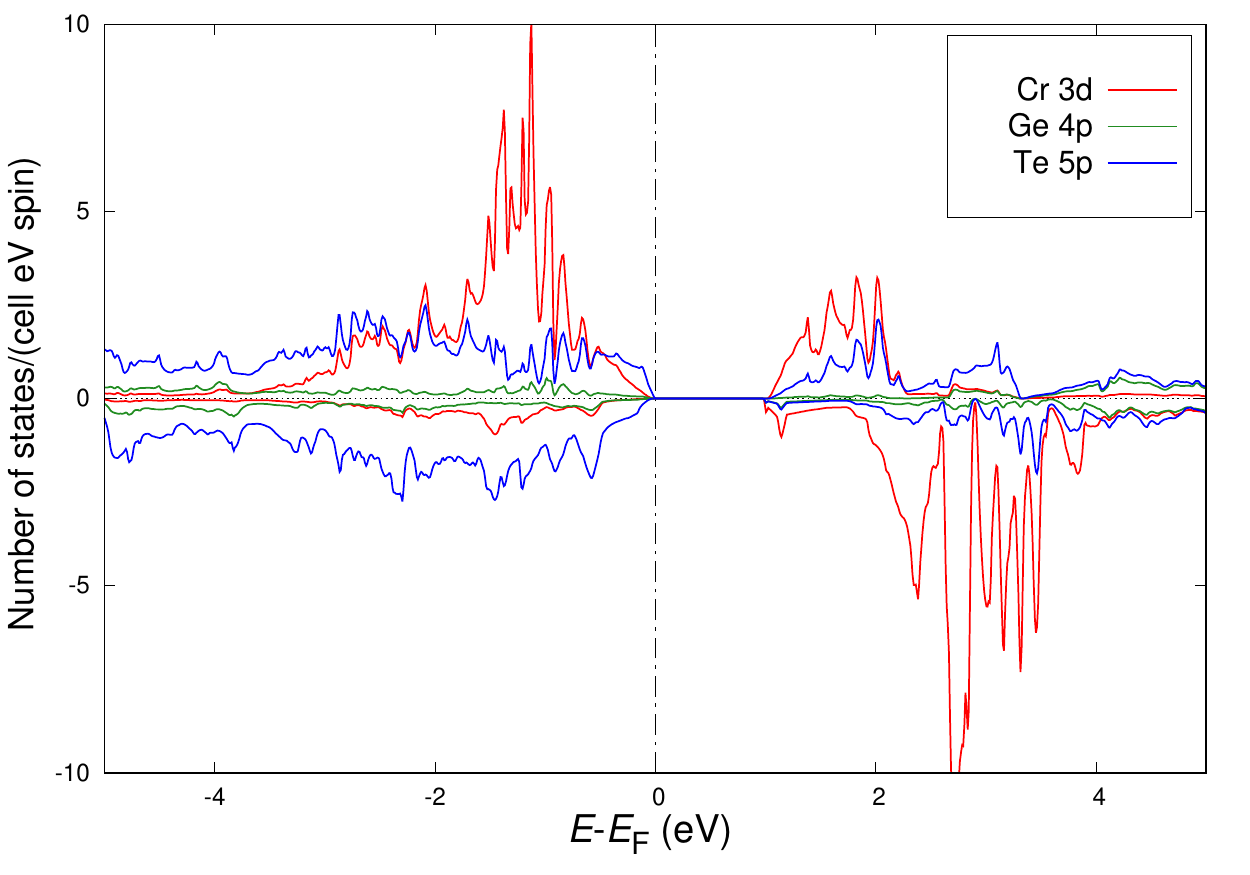}
	\caption{The partial density of states projected on Cr-$3d$, Ge-$4p$, and Te-$5p$ states in CrGeTe$_{3}$ calculated in DFT-PBE (top) and QSGW (bottom).}
	\label{fig:Cgt3pdos}	
\end{figure}

\clearpage
\subsubsection{Fe$_{3}$GeTe$_{2}$}
\begin{figure}[hbt]
	\centering
	\includegraphics[width=1.0\linewidth]{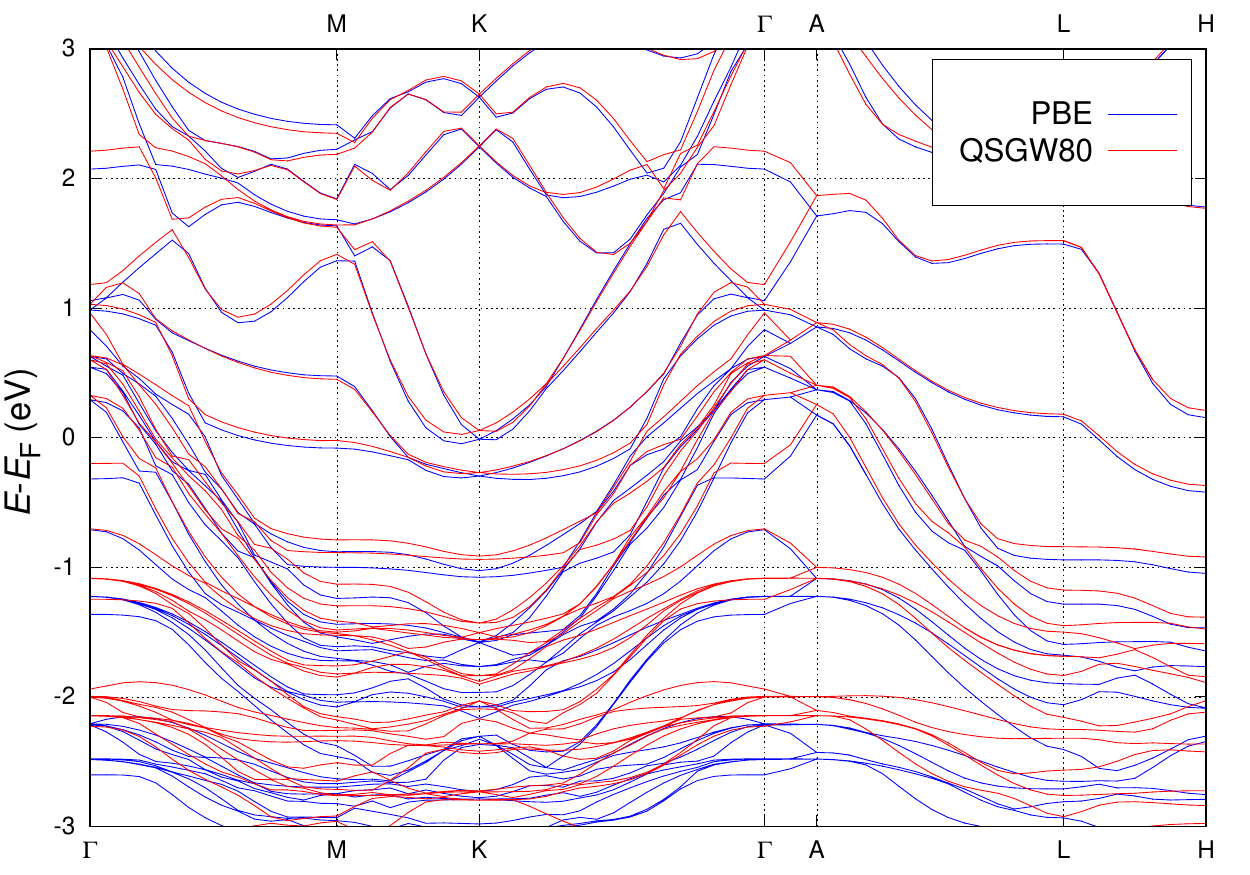} \\
	\includegraphics[width=1.0\linewidth]{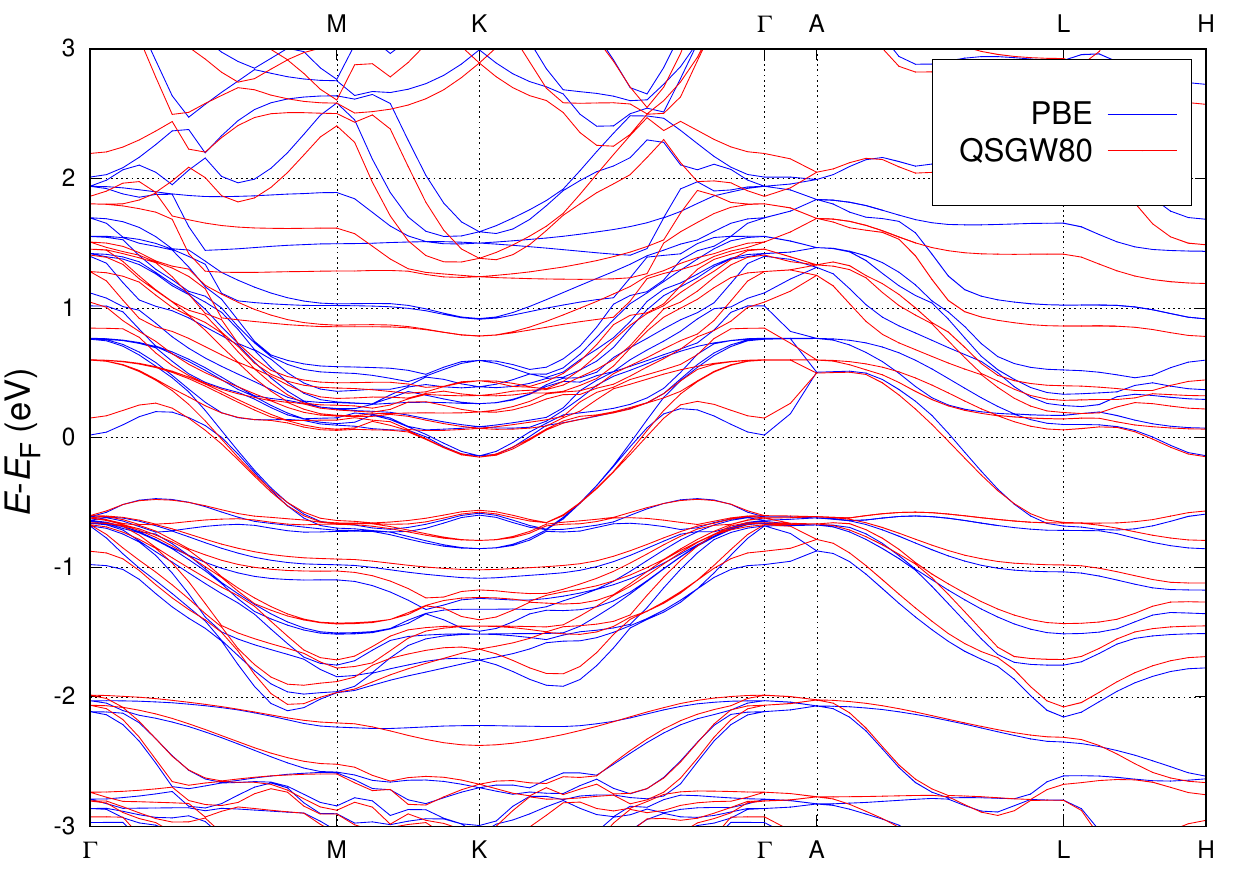}
	\caption{Spin-polarized band structures of Fe$_{3}$GeTe$_{2}$ calculated in DFT and QSGW.
          The top (bottom) panel shows the majority (minority) spin state.}
	\label{fig:fgt1band}	
\end{figure}
\begin{figure}[hbt]
  \centering
  \vspace*{42pt}  
	\includegraphics[width=1.0\linewidth]{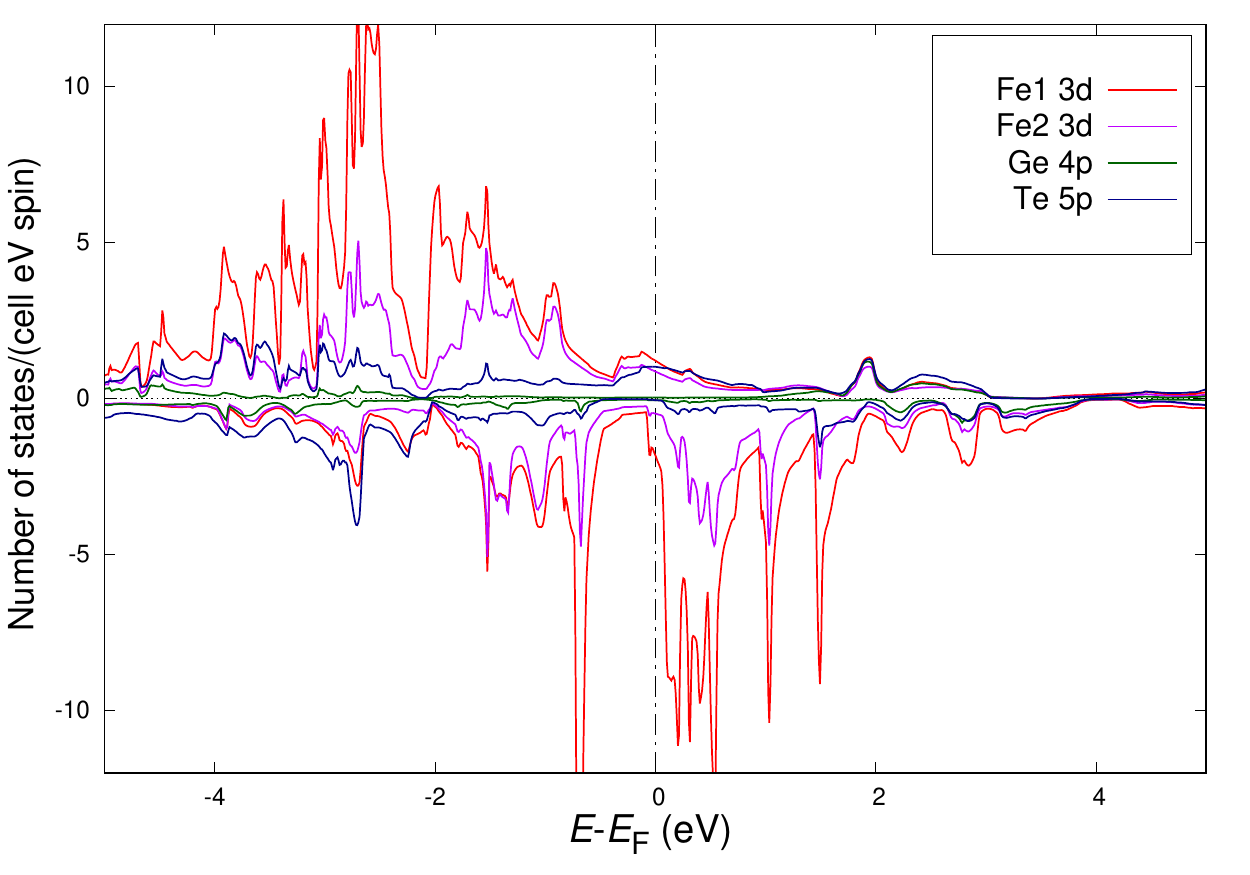} \\
	\includegraphics[width=1.0\linewidth]{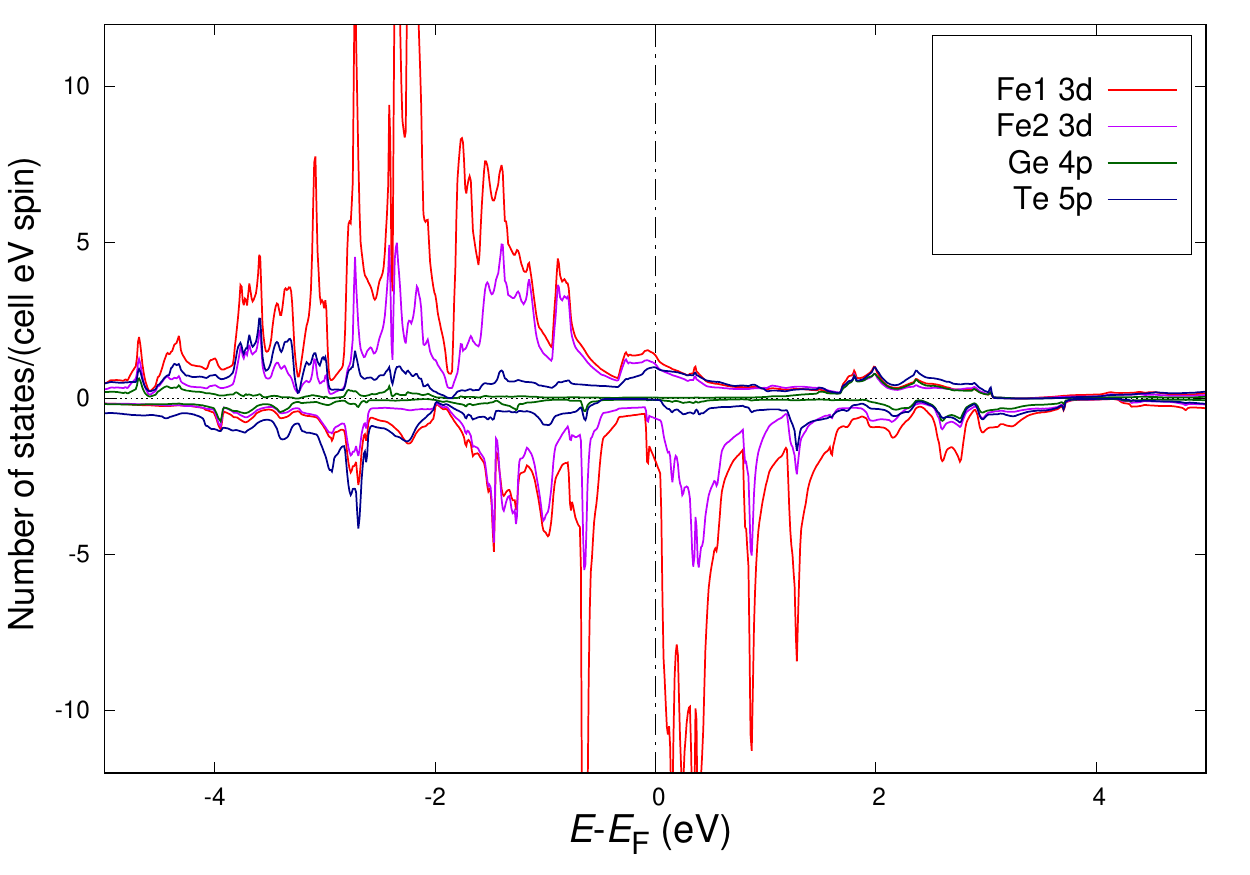}
	\caption{The partial density of states projected on Fe-$3d$, Ge-$4p$, and Te-$5p$ states in Fe$_{3}$GeTe$_{2}$ calculated
           in DFT (top) and QSGW (bottom).}
	\label{fig:fgt1dos}	
\end{figure}

\clearpage

\subsubsection{CrI$_{3}$}

\begin{figure}[htb]
	\centering
	\includegraphics[width=1.0\linewidth]{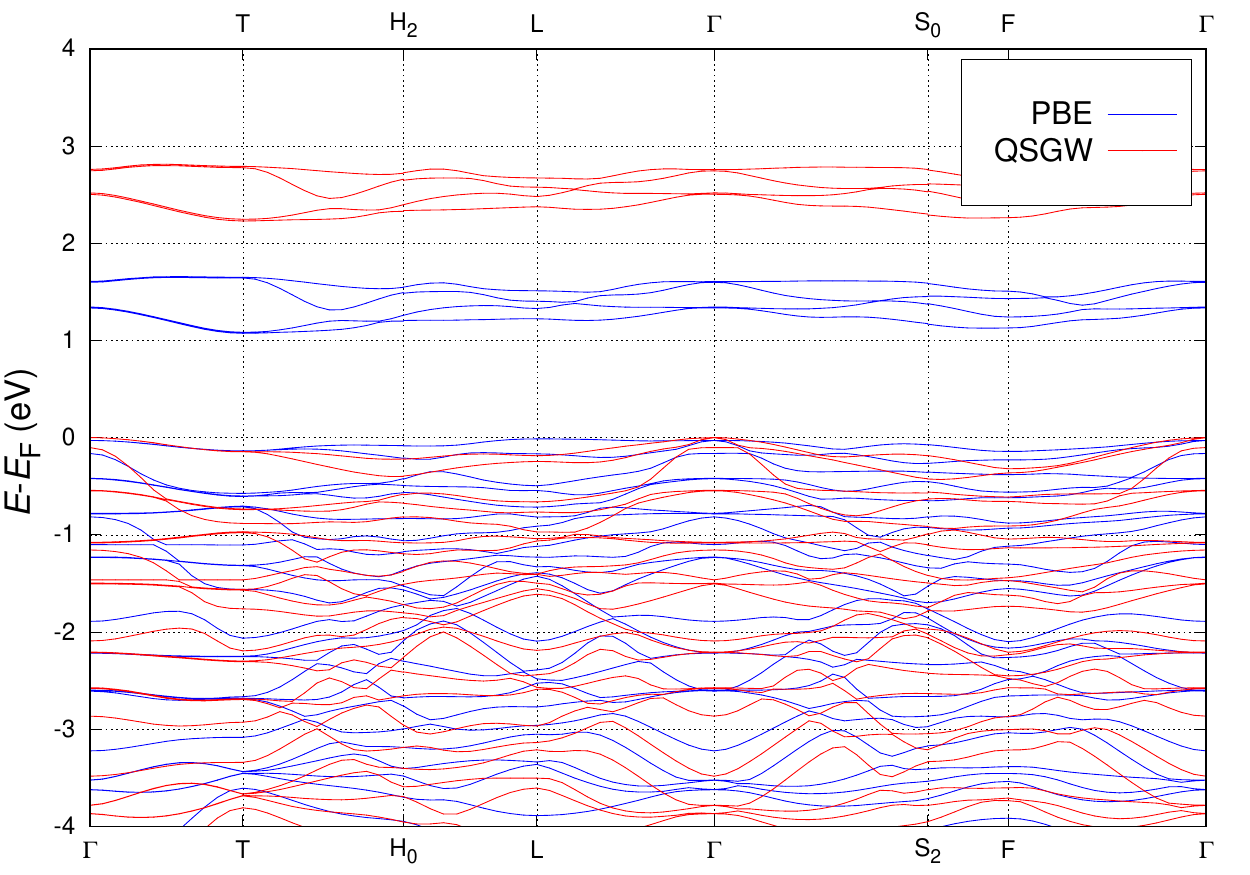} \\
	\includegraphics[width=1.0\linewidth]{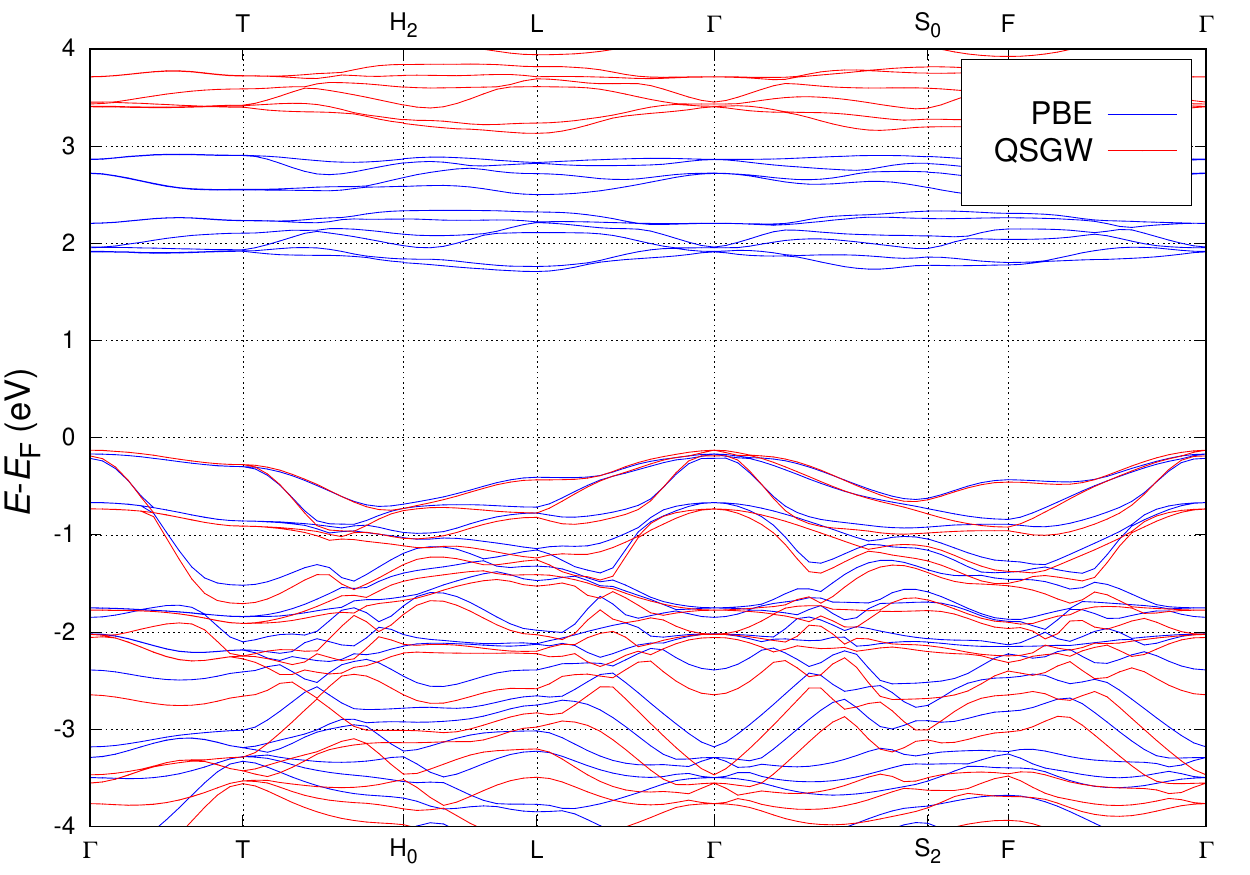} 
	\caption{Spin-polarized band structures of CrI$_{3}$ calculated in DFT and QSGW.
          The top (bottom) panel shows the majority (minority) spin state.
          QSGW shifts the unoccupied states up and increases the bandgap.} 
\end{figure}

\newpage

\subsection{DOS calculated in DFT$+U$, DFT, and QSGW}
\subsubsection{CrI$_{3}$}
\begin{figure}[htb]
  	\begin{tabular}{c}
	\includegraphics[width=1.0\linewidth]{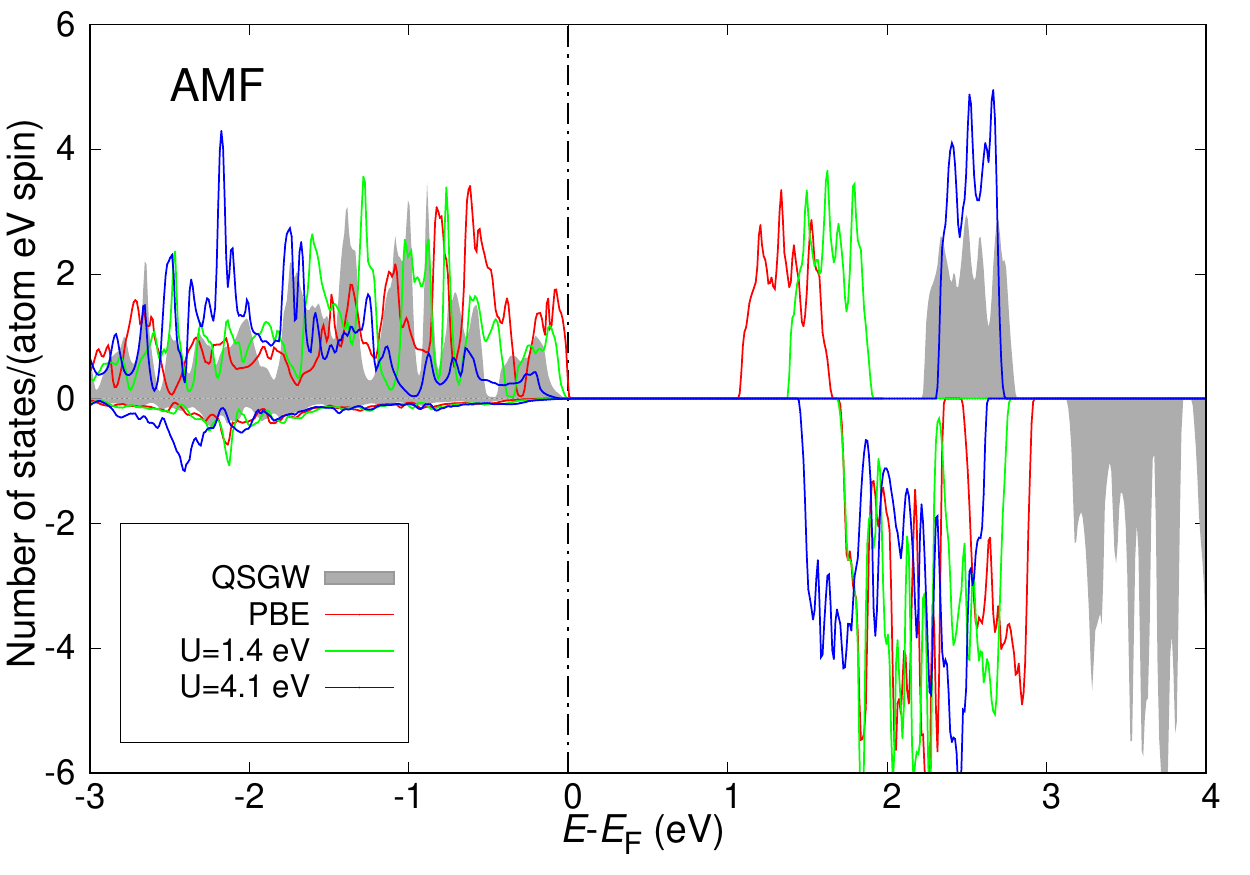}\\
	\includegraphics[width=1.0\linewidth]{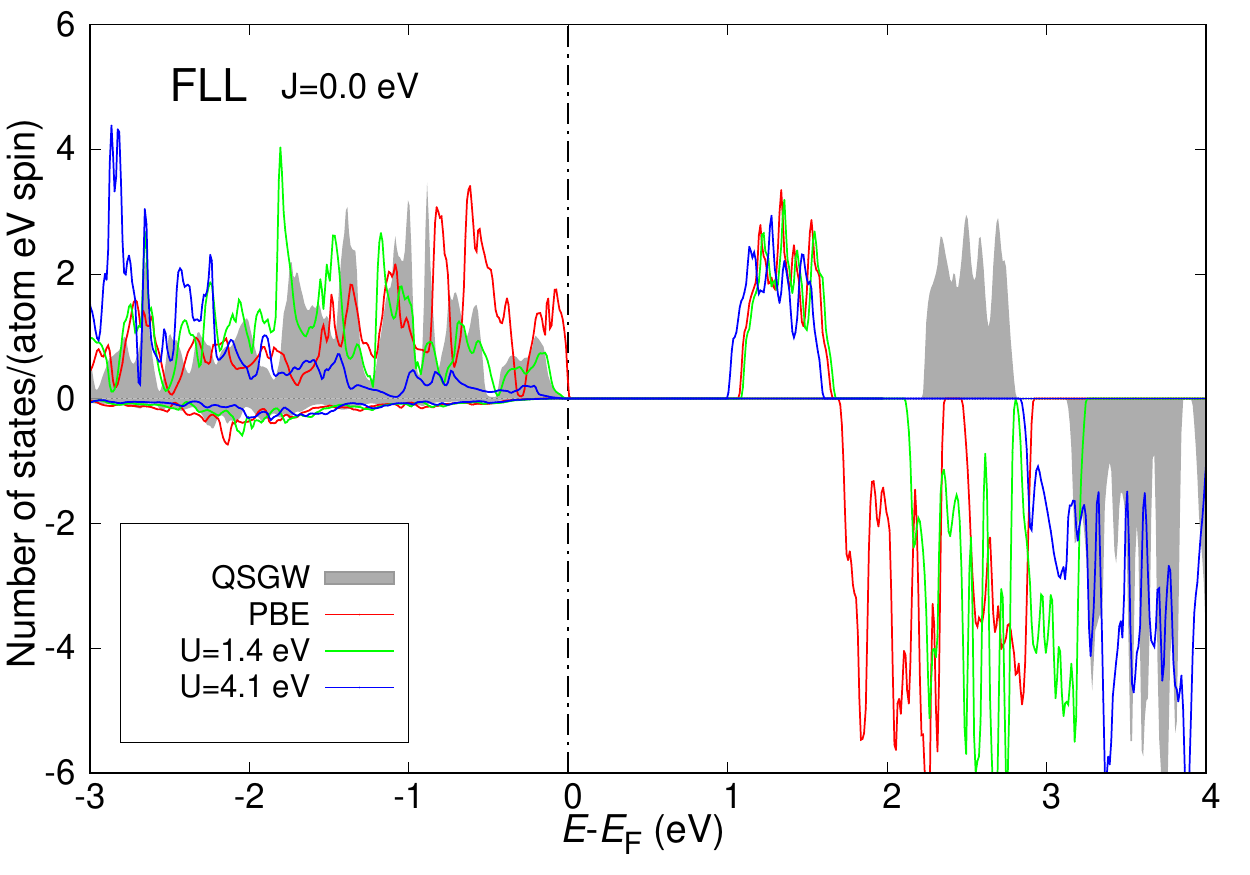}
	\end{tabular}        
        \caption{The partial density of states projected on Cr-$3d$ states in CrI$_{3}$.
          Both AMF (top) and FLL (bottom) double-counting schemes are employed for the DFT$+U$ calculations.}
\end{figure}

\clearpage

\subsubsection{CrGeTe$_{3}$}
\begin{figure}[hbt]
	\centering
	\includegraphics[width=1.0\linewidth]{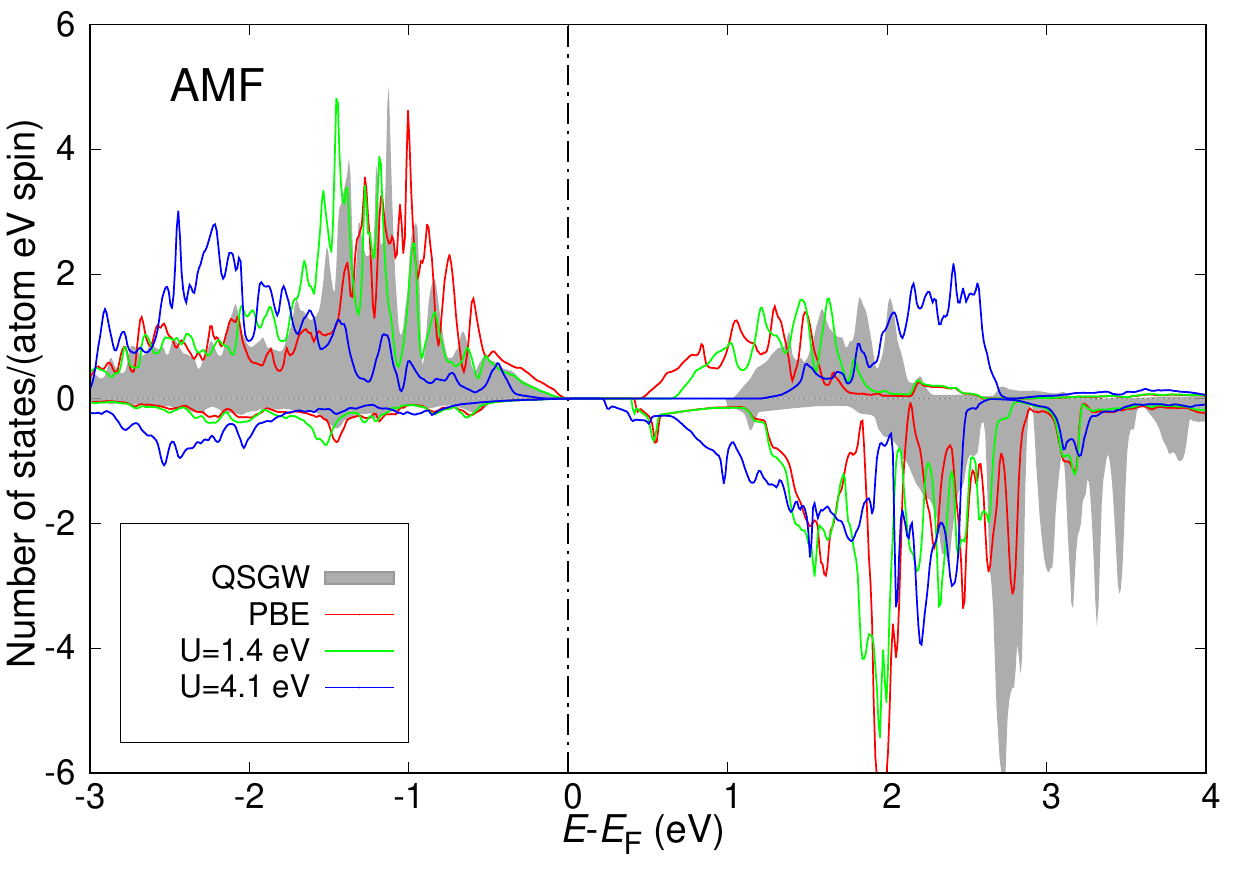}\\
	\includegraphics[width=1.0\linewidth]{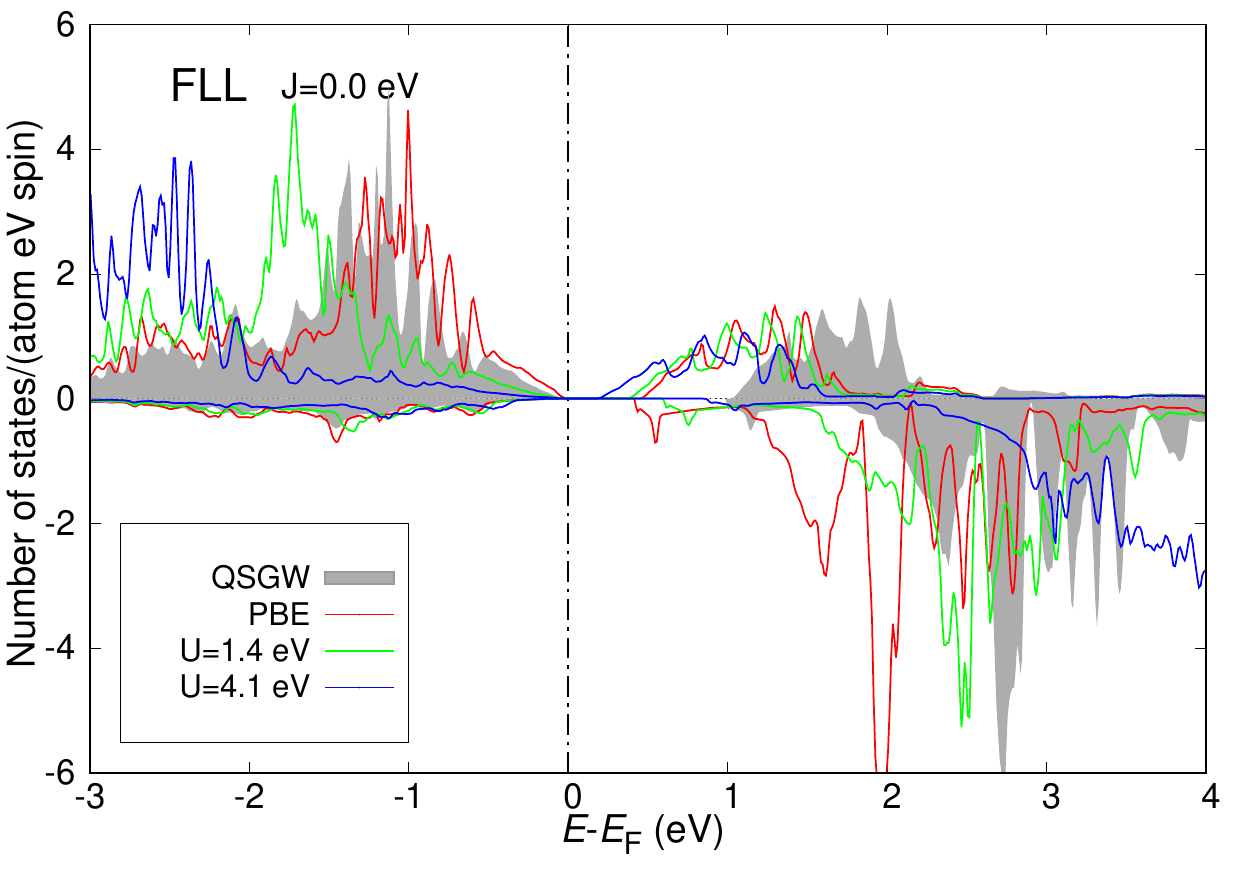}
        \caption{The partial density of states projected on Cr-$3d$ states in CrGeTe$_{3}$.
          Both AMF (top) and FLL (bottom) double-counting schemes are employed for the DFT$+U$ calculations.          
}
\end{figure}


\subsubsection{VI$_{3}$}
\begin{figure}[hbt]
	\centering
	\includegraphics[width=1.0\linewidth]{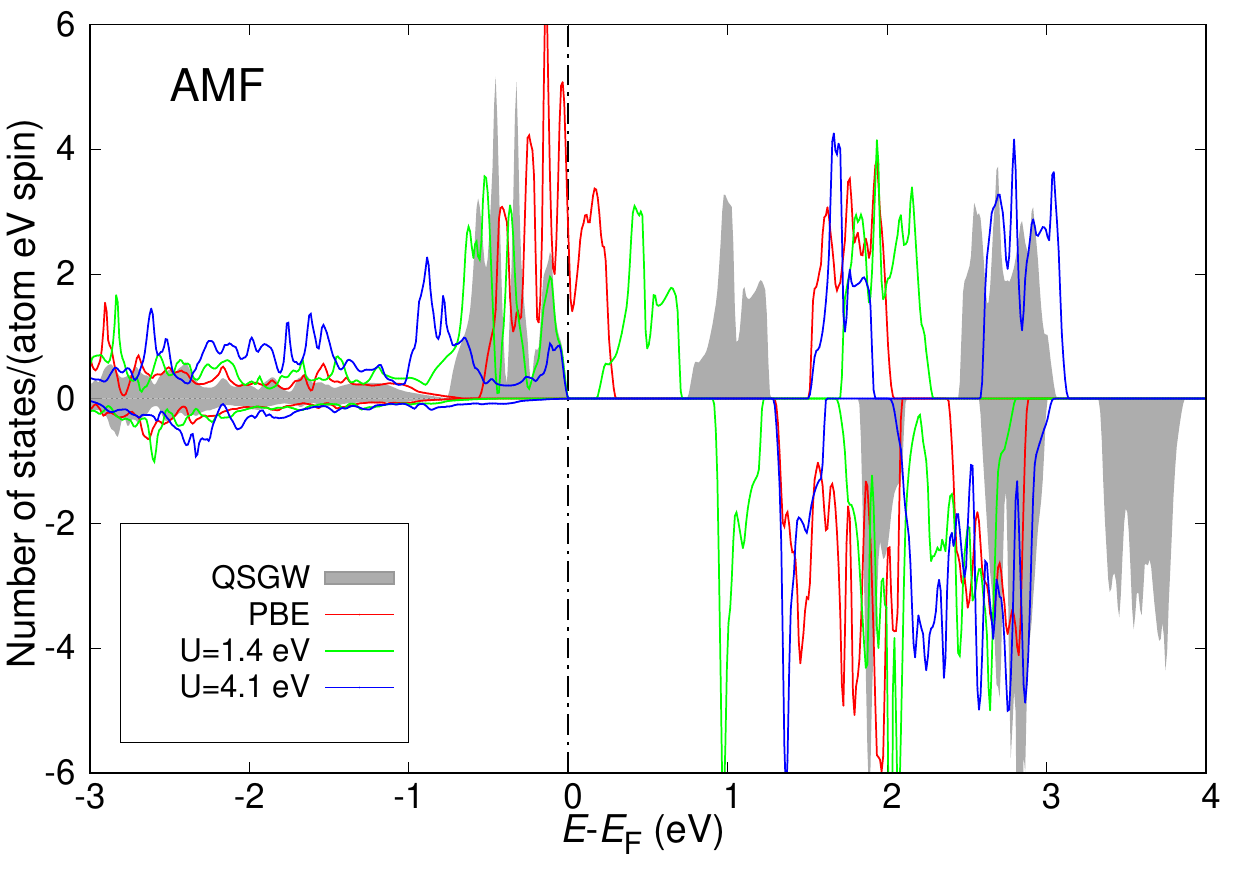}\\
	\includegraphics[width=1.0\linewidth]{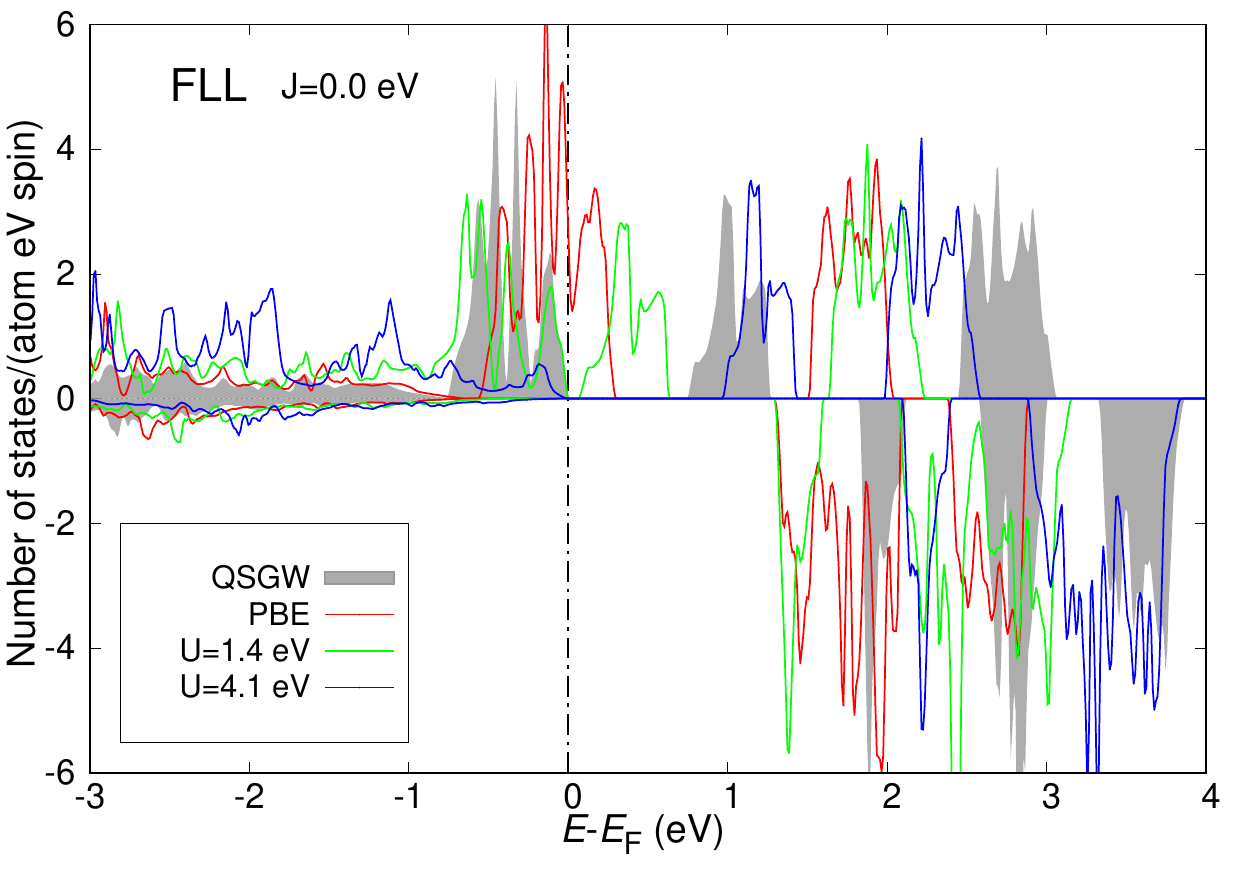}\\
	\includegraphics[width=1.0\linewidth]{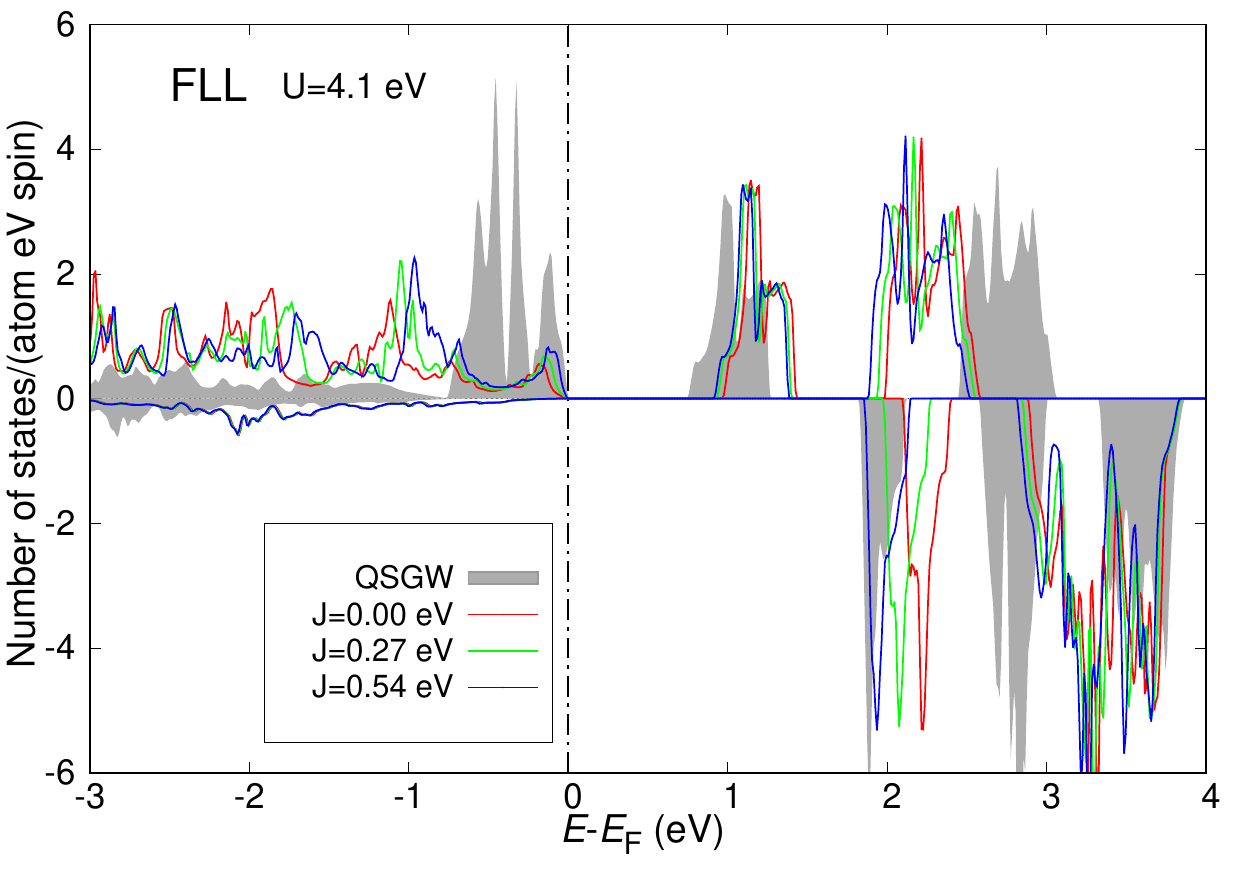}
	\caption{The partial density of states projected on V-$3d$ states.
          The top panel shows DFT$+U$ results calculated using the AMF scheme.
          DFT$+U$ calculations in the middle and bottom panel are carried out using the FLL scheme.
          The middle panel is calculated using various $U$ values and $J=$\SI{0}{\eV}.
          The bottom panel is calculated using various $J$ values and $U=$\SI{4.1}{\eV}.          
}
\end{figure}

\clearpage

\end{document}